\documentclass[twocolumn,showpacs,preprintnumbers,amsmath,amssymb]{revtex4}

\usepackage{graphicx}
\usepackage{dcolumn}
\usepackage{bm}
\usepackage{amsfonts}
\newtheorem{guess5}{Proposition}[section]
\newtheorem{guess}{Theorem}[section]
\newtheorem{guess1}{Note}[section]
\newcommand{\nc}{\newcommand}
\nc{\nt}{\indent}
\nc{\be}{\begin{equation}}
\nc{\ee}{\end{equation}}
\nc{\bd}{\begin{displaymath}}
\nc{\ed}{\end{displaymath}}
\nc{\bi}{\begin{itemize}}
\nc{\ei}{\end{itemize}}
\nc{\bnu}{\begin{enumerate}}
\nc{\enu}{\end{enumerate}}
\nc{\bde}{\begin{definition}}
\nc{\ede}{\end{definition}}
\nc{\bn}{\begin{guess1}}
\nc{\en}{\end{guess1}}
\nc{\bes}{\begin{guess2}}
\nc{\ees}{\end{guess2}}
\nc{\bt}{\begin{guess}}
\nc{\et}{\end{guess}}
\nc{\bpb}{\begin{guess4}}
\nc{\epb}{\end{guess4}}
\nc{\bp}{\begin{guess5}}
\nc{\ep}{\end{guess5}}
\nc{\bco}{\begin{guess6}}
\nc{\eco}{\end{guess6}}
\nc{\bl}{\begin{guess7}}
\nc{\el}{\end{guess7}}
\nc{\C}{\mathbb{C}}
\nc{\R}{\mathbb{R}}
\nc{\Rt}{\mathbb{R^{\rm 3}}}
\nc{\Cn}{\mathbb{C^{\it n}}}
\nc{\Cm}{\mathbb{C^{\it m}}}
\nc{\Rn}{\mathbb{R^{\it n}}}
\nc{\Rnn}{\mathbb{R^{\it 2n}}}
\nc{\Rm}{\mathbb{R^{\it m}}}
\nc{\dm}{{\bf Dimostrazione.} \hspace{0.1cm}}
\nc{\brm}{\begin{rm}}
\nc{\erm}{\end{rm}}
\nc{\br}{\langle}
\nc{\ke}{\rangle}
\nc{\q}{\forall}
\nc{\bc}{\begin{center}}
\nc{\ec}{\end{center}}
\nc{\blt}{\left (}
\nc{\brt}{\right )}
\nc{\blmo}{\left |}
\nc{\brmo}{\right |}
\nc{\brq}{\right ]}
\nc{\blq}{\left [}
\nc{\brg}{\right \}}
\nc{\blg}{\left \{}
\nc{\h}{\hat}
\nc{\s}{\overline}
\nc{\ch}{\check}
\nc{\la}{\lambda}
\nc{\al}{\alpha}
\nc{\om}{\omega}
\nc{\Om}{\Omega}
\nc{\ii}{\infty}
\nc{\os}{\hspace{0.1cm}}
\nc{\vs}{\vspace{0.1cm}}
\nc{\lni}{\lim_{n \to \infty}}
\nc{\lki}{\lim_{k \to \infty}}
\nc{\sui}{\sum_{i=1}^{\infty}}
\nc{\sni}{\sum_{n=1}^{\infty}}
\nc{\suk}{\sum_{k=1}^{n}}
\nc{\sun}{\sum_{i=1}^{n}}
\nc{\sik}{\sum_{i=1}^{k}}
\nc{\sjn}{\sum_{j=1}^{n}}
\nc{\iin}{i=1,2,\dots,n}
\nc{\oos}{\hspace{0.2cm}}
\nc{\ooos}{\hspace{0.6cm}}
\nc{\vvs}{\vspace{0.2cm}}
\nc{\vvvs}{\vspace{1cm}}
\nc{\lmt}{\longmapsto}
\nc{\D}{\mathcal{D}}
\nc{\CC}{\mathcal{C}}
\nc{\ES}{\mathcal{S}}
\nc{\EE}{\mathcal{E}}
\nc{\eps}{\varepsilon}
\nc{\ff}{\varphi}
\nc{\mii}{\int_{-\infty}^{\infty}}
\nc{\mpp}{\int_{-\pi}^{\pi}}
\nc{\iab}{\int_{a}^{b}}
\nc{\tf}{\hat{f}}
\nc{\hy}{\hyphenation}
\nc{\bna}{\boldsymbol{\nabla}}
\nc{\bon}{\boldsymbol{n}}
\nc{\xz}{x_{0}}
\nc{\yz}{y_{0}}
\nc{\tz}{t_{0}}
\nc{\tl}{\widetilde}
\nc{\lb}{\label}
\nc{\frsq}{\begin{flushright}
$\square$
\end{flushright}
}

\begin{document}

\title{Classical billiards and double-slit quantum interference}

\author{Giacomo Fonte}

\affiliation{Dipartimento di Fisica e Astronomia, Via S. Sofia 64, I-95123 Catania, Italy and Istituto Nazionale di Fisica Nucleare, Sezione di Catania, Via S. Sofia 64, I-95123 Catania, Italy}
\email{giacomo.fonte@ct.infn.it}
\author{Bruno Zerbo}
\affiliation{Dipartimento di Fisica e Astronomia, Via S. Sofia 64, I-95123 Catania, Italy and Istituto Nazionale di Fisica Nucleare, Sezione di Catania, Via S. Sofia 64, I-95123 Catania, Italy }

\date{October 21, 2011}

\begin{abstract}

\ooos We carry out a numerical simulation   about  the occurrence of  interference fringes  in experiments  where an initial Gaussian wave packet evolves inside  a billiard domain   with two  slits on the boundary. Our simulation  extends  a previous work by  Casati and Prosen and it is aimed to test their surprising conclusion that the fringes disappear in the experiments with fully chaotic billiards. According to the results found,  we are led  to reassess this remarkable  effect of classical dynamics on  quantum interference. Actually, we highlight another factor  which  acts on  interference: a symmetry condition $(SC)$ concerning  the experimental set-up. This condition seems even to play  a  role more important than classical chaos. Indeed, when the $SC$ is verified, classical chaos has no effect, whereas when the $SC$ is violated classical chaos turns out to be an additional factor that causes dephasing at the slits. We explain the respective roles of these two factors, by specifying the physical mechanism through which they influence the interference patterns. This mechanism depends both on the position and direction of the initial wave packet  and on certain  its recurrences which occur especially in the regular billiards.\\

\end{abstract}

\maketitle                            

\section{Introduction}
 \ooos In a recent  paper \cite{ca}, Casati and Prosen have shown an interesting numerical simulation concerning the quantum-mechanical  time-evolution of an initial Gaussian wave packet inside a billiard  domain having the following shapes:
\bi
\item [$s_{1})$] isosceles right triangle,
\item [$s_{2})$] as above but with the hypotenuse replaced by a circular arc.
\ei
In both cases the domain has two narrow slits on a cathetus, through which  the  probability current leaks out bit by bit. This current is integrated in time on a screen placed at a certain distance from the cathetus with the slits, until the probability  inside the billiard becomes vanishingly small. The set-up of the experiment is intriguing, because it involves properties of classical dynamics and properties of  quantum dynamics. The former are represented by the shapes of the billiards, where  classical  dynamics can be regular or chaotic, whereas the latter are represented by the interference fringes in the time-integrated probability current. The experiment provides thus  a new  opportunity to investigate a question, which although discussed for a long time, still remains far from being fully understood, {\em i.e.}  the manifestations of  classical chaos in quantum mechanics (quantum chaos) \cite {qc}.  Casati and Prosen  found clear interference fringes  in the case of the regular billiard $s_{1})$ but no sign of interference in the case of the chaotic billiard  $s_{2})$. They traced this surprising fact back to  the randomization due to the chaotic nature of the classical dynamics inside the billiard $s_{2})$ but they did not specify any physical mechanism that could explain in detail the phenomenon. Quite recently, Levnaji\'{c} and Prosen have carried out \cite{pro} a variant of the numerical experiment  in \cite{ca} and they have  confirmed that  classical chaos is the fundamental cause of the disappearance of the fringes. The  Casati-Prosen work has also stimulated real laboratory experiments. Tang {\em et al.} \cite{tan} have performed double-slit experiments with water surface waves. Bittner {\em et al.} \cite{dietz} have instead performed double-slit experiments with microwave billiards. The experiment \cite{tan} does not reproduce all the features of the numerical simulation \cite{ca}, but it equally shows that classical chaos has a certain effect on interference. The experiment  \cite{dietz} is quite similar to the simulation \cite{ca}, because it employs directional wave packets and displays  time-averaged intensity patterns, but the authors are somewhat critical of the role of classical chaos. Indeed, they show that the results in \cite{ca} can be reproduced only approximately for certain positions and directions of the initial wave packet. Moreover, according to an unpublished preliminary version \cite{fon} of the present paper, they  also show that the symmetry of the experimental set-up  plays a certain role. In the end, they  conclude that further investigations are necessary for a thorough understanding  of this kind of double-slit experiments. We agree with the authors of \cite{dietz} on this point and also note that after all the original investigation \cite{ca} is based on only two numerical simulations and that the recent variant  \cite{pro} of it  deals again with triangular billiards. For all these reasons, we present here a numerical investigation that is both a revision  of  our previous work  \cite{fon} and an extension   to a larger set of experiments  where we vary the center $P_{0}$ of the initial Gaussian wave packet and the direction of its wave vector $\boldsymbol{k}$. As matter of fact, our numerical outcomes turn out to be in agreement with the experimental ones in \cite{dietz} and they  confirm and clarify the role played by the symmetry of the experimental set-up. This symmetry seems even to play a role  more important than classical chaos. Indeed, when it is verified classical chaos has no effect,  because, irrespective of the classical  integrability of the billiards, we always find symmetric interference patterns with a fringe visibility of $100\%$.  When it is violated, we find instead a complex scenario of outcomes  where the fringes can be  either practically absent or present with a visibility that varies greatly, even in the experiments with regular billiards. In practice, each experiment has its own outcome, according to the specific violation of the symmetry   and according to the classical integrability of the  billiard. We will give  a qualitative explanation for most of our experiments, by specifying  a likely physical mechanism which depends both on the  violation of the symmetry from  the initial state and on the classical integrability of the billiards.\\

 The paper is organized as follows. In Sec. II, we describe our numerical experiments. In Sec. III, we highlight the role of the symmetry of the experimental set-up. In Sec.IV, we investigate the effects of the violation of this symmetry on the part of the initial state and the effects due to the classical integrability of the billiards.

\section{Description of the numerical experiment}

\ooos The quantum dynamics of  an initial wave packet $\psi_{0}$ inside a  billiard $B$ is given by the solution of the initial/boundary-value problem \footnote{Atomic units   are used throughout the paper. We set $m=1$.}
\be\lb{eprmi}
\left\{ \begin{array}{ll}
i \frac{\partial}{\partial t}\psi (x,y;t)=-\frac{1}{2}\triangle \psi(x,y;t)\\
 \psi (x,y;0)=\psi_{0} (x,y)\\
\psi(x,y;t)\arrowvert_{\partial B} =0, \q \;t \geq 0,
  \end{array} \right.
\ee
where  $\partial B$ denotes the billiard boundary.
\begin{figure}
\begin{center}
\includegraphics[scale=0.6]{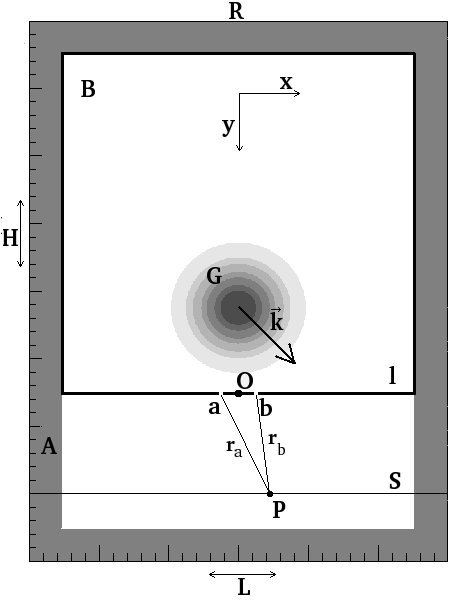}
\end{center}
\caption{Experiment set-up. $H$  height and  $L$ length of the integration region $R$,
 $A$ absorbing layer, $B$ billiard, represented by  a square, $l$ side of $B$, $O$ origin of the coordinate system, $a$ and $b$ slits, $S$ viewing screen, $G$  initial Gaussian wave packet (\ref{egwp}),  $\boldsymbol{k}=(k_{x},k_{y})$  wave vector of $G$.}
 \lb{setup}
\end{figure}

This problem has been solved  in \cite{ca} by a finite difference  method  in the  case of the triangular billiards   $ s_{1}), s_{2})$ and   with $\psi_{0}(x,y)$  given by the normalized Gaussian wave packet
\be\lb{egwp}
\psi_{0}(x,y)=(2\pi \sigma_{0}^{2})^{-1/2}e^{i(k_{x}x+k_{y}y)}e^{-[(x-x_{0})^{2}+(y-y_{0})^{2}]/4\sigma_{0}^{2}}.
 \ee

   Owing to the simple geometry of the billiards $ s_{1}), s_{2})$, the   boundary condition $\psi(x,y;t)\arrowvert_{\partial B} =0$   was handled  in \cite{ca} easily.  Since it is our intention here  to examine billiards of various shapes, we find it more convenient to force the boundary condition by means of a narrow smooth potential barrier $V_{B}(x,y)$, greater than zero along the borders of the billiard  (but equal to zero at the two slits) and sufficiently high   to reduce as much as possible the tunnel effect. Thus, instead of the  initial/boundary-value problem (\ref{eprmi}), we solve the Cauchy initial-value problem in the whole space $\mathbb{R^{\rm 2}}$

\be\lb{eptch}
\left\{ \begin{array}{ll}
i \frac{\partial}{\partial t}\psi (x,y;t)=H\psi(x,y;t)\\
\psi (x,y;0)=\psi_{0} (x,y),
  \end{array} \right.
\ee
where
\be\lb{eha}
H=-\frac{1}{2}\triangle + V_{B}(x,y),
\ee
and $\psi_{0} (x,y)$ is still the wave packet (\ref{egwp}). The replacement of the problem (\ref{eprmi}) with the problem  (\ref{eptch}) has also  formal advantages. Indeed,  contrary to  (\ref{eprmi}),  the problem (\ref{eptch}) remains the same whatever the billiard shape, {\em i.e.} it is always defined  in the whole space $\mathbb{R^{\rm 2}}$ and presents a Hamiltonian operator (\ref{eha}) which turns out to be  always (\cite{si}, p. 184) self-adjoint in the space $L^{2}(\mathbb{R^{\rm 2}})$. This latter fact is essential so that (\ref{eptch}) can admit in $L^{2}(\mathbb{R^{\rm 2}})$ (\cite{sii}, p. 105) a unique solution. This is formally written as
\be\lb{eforsol}
\psi(x,y;t)= \int_{-\ii}^{\ii}e^{-i\la t}dE(\la)\psi_{0} (x,y),
\ee
where  $\{E(\la)\}$ is the spectral family (\cite{sii}, p. 24) of $H$.\\

     The replacement of (\ref{eprmi}) with  (\ref{eptch}) is the main technical difference of the present numerical simulation in comparison with that in \cite{ca}, whereas the integration of  (\ref{eptch}) has been carried out  following a method roughly  similar to that in \cite{ca}. We approximate the exact solution (\ref{eforsol}), by expanding the exponential up to the fourth order, {\em i.e.}
 \bd
 \psi(x,y;t) \simeq  \int_{-\ii}^{\ii}\sum _{n=0}^{4} \frac{1}{n!}(-i\la t)^{n}dE(\la)\psi_{0} (x,y)=
 \ed
 \be\lb{eper}
 =\sum _{n=0}^{4} \frac{1}{n!}(-i H t)^{n}\psi_{0} (x,y),
 \ee
 where we have used
 \bd
 \int_{-\ii}^{\ii}\la^{n}dE(\la) =H^{n}, n=0,1,2,3,4.
 \ed

 Then we  introduce a grid
 \bd
 x_{m}=m\delta, y_{n}=n\delta, t_{p}=p\tau,
 \ed
 \bd
 m,n=0,\pm 1,\pm 2,\dots, p=0,1,2,\dots,
 \ed
 of discrete values for $x,y,t$, to evaluate each term in (\ref{eper}) through standard difference formulas. In this way we get  an explicit differencing scheme, {\em i.e.} a finite difference equation where the values of (\ref{eper}) at each node of the spatial grid $R$  (see the experimental set-up in Fig. \ref{setup}) at time $(p+1)\tau$ are given directly in terms of the corresponding values at the earlier time $p\tau$. In order to have numerical convergence within machine precision, we employ the value $2\cdot 10^{-3}$ for the space step $\delta$ and the value $10^{-6}$ for the time step $\tau$. Notice that with these values for $\delta$ and $\tau$,  truncations of order lower than the fourth  give rise (see also Ref. \cite{ma} and references therein) to serious problems of stability and probability conservation. Strictly speaking the integration region of (\ref{eptch}) is the whole space $\mathbb{R^{\rm 2}}$, but in our calculations we have reduced this space to the finite rectangular  region $R$. This reduction causes reflections from the border of $R$. In order to avoid this problem, we have employed the same method in \cite{ca}, that is  damping the wave function by means of a smooth imaginary potential $-iV_{A}(x,y)$, where $V_{A}(x,y) > 0$ inside an ``absorbing layer'' $A$  along the border of $R$ (see Fig. 1) and zero elsewhere. The integration of (\ref{eptch}) with the method just described turns out to be very accurate, the total probability is conserved and  its  transmission through  the boundary of $B$ and its reflection from the sides of $R$ are negligible. The intensity $I(x)$ on the screen  $S$ at $x$, {\em i.e.} the probability density for a particle to be detected at the point $x$ of $S$ at the end of the experiment,  is then evaluated by
\be\lb{etini}
I(x)= \int_{0}^{T}j_{y}(x,\bar{y};t)dt,
\ee
where $T$ is the duration of the experiment, $j_{y}$ the component along the $y$-axis of the probability  current and $\bar{y}$ the position of the screen.
\begin{table}
\begin{center}
\begin{tabular}{|l|r|}
\hline
Parameter & Value \\
\hline
$H$ (height of the integration region)     & $1.6$ \\
\hline
$L$ (length of the integration region)   &  $1.2$ \\
\hline
width of the absorbing layer $A$ & $0.1$ \\
\hline
 $l$ (length of the billiard basis)  & $1$ \\
\hline
 width of the potential barrier $V_{B}(x,y)$ & $0,008$ \\
\hline
 height of the potential barrier $V_{B}(x,y)$ & $10^{6}$ \\
\hline
$w$ (width of the slits) & $0.012$ \\
\hline
$d$ (distance between the slits) & $0.1$ \\
\hline
$s$ (distance of the screen from billiard basis) & $0.3$\\
\hline
$\delta$ (space step size) & $0.002$\\
\hline
$\tau$ (time step size) & $10^{-6}$\\
\hline
$\sigma_{0}$ (rms width of the packet (\ref{egwp})) & $0.09$\\
\hline
$\|\boldsymbol{k}\| =\sqrt{k_{x}^{2}+k_{y}^{2}}$ (length of the wave vector $\boldsymbol{k}$) & $180$ \\
\hline
\end{tabular}
\end{center}
\caption{Parameters employed in our numerical simulation. Values in atomic units.}
\lb{tab1}
\end{table}

As regards the parameters, we have chosen (see Table \ref{tab1}) values deliberately close or equal to those in \cite{ca}.

\section{The role of the symmetry of the experimental set-up}
In this section  we focus  on the role of the symmetry of the experimental set-up in our interference experiments.\\

\begin{figure*}
\begin{center}
\begin{tabular}{|c|c|c|}
\hline
Column I & Column II & Column III\\
\hline
a) \includegraphics[scale=0.171]{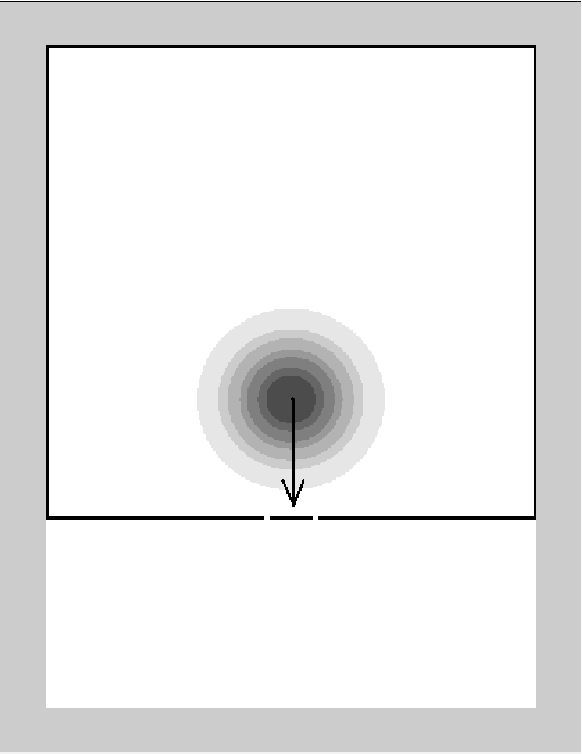} & \includegraphics[height=3.2cm]{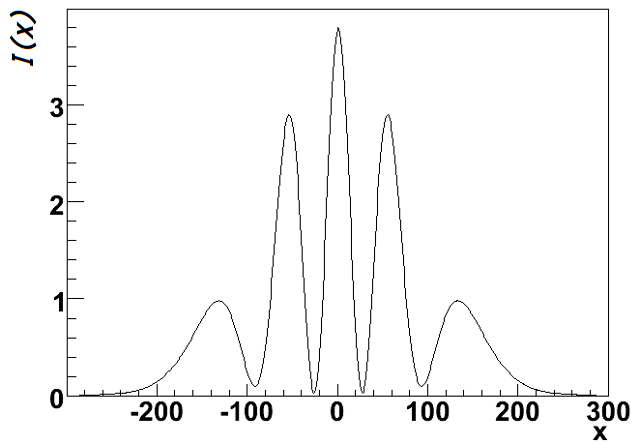} &         \includegraphics[height=3.2cm]{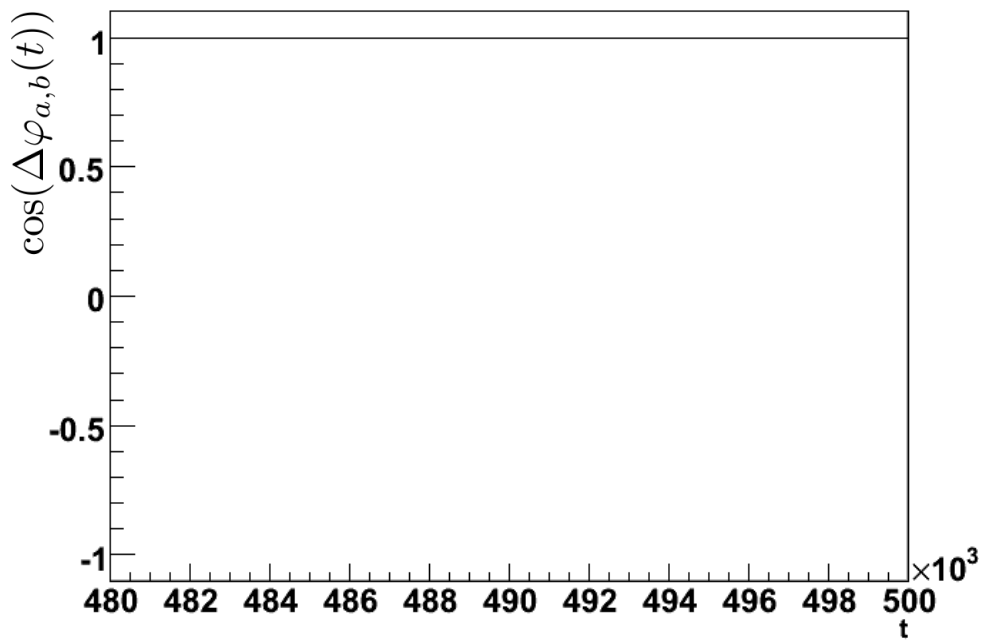} \\
\hline
b) \includegraphics[scale=0.171]{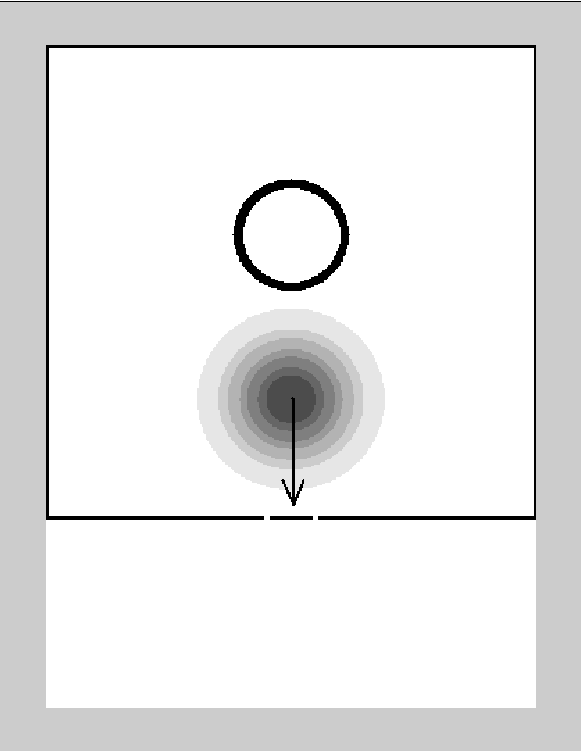} & \includegraphics[height=3.2cm]{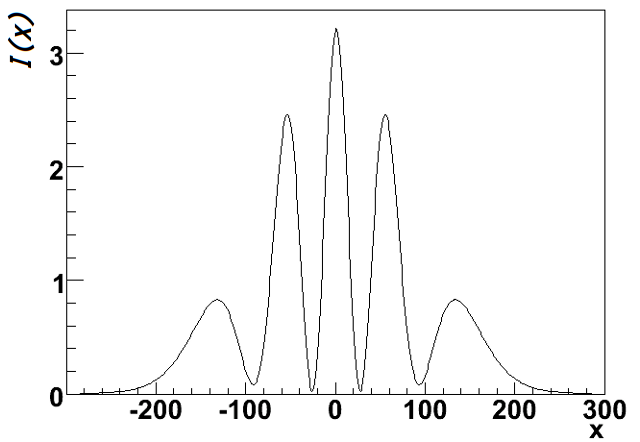} &       \includegraphics[height=3.2cm]{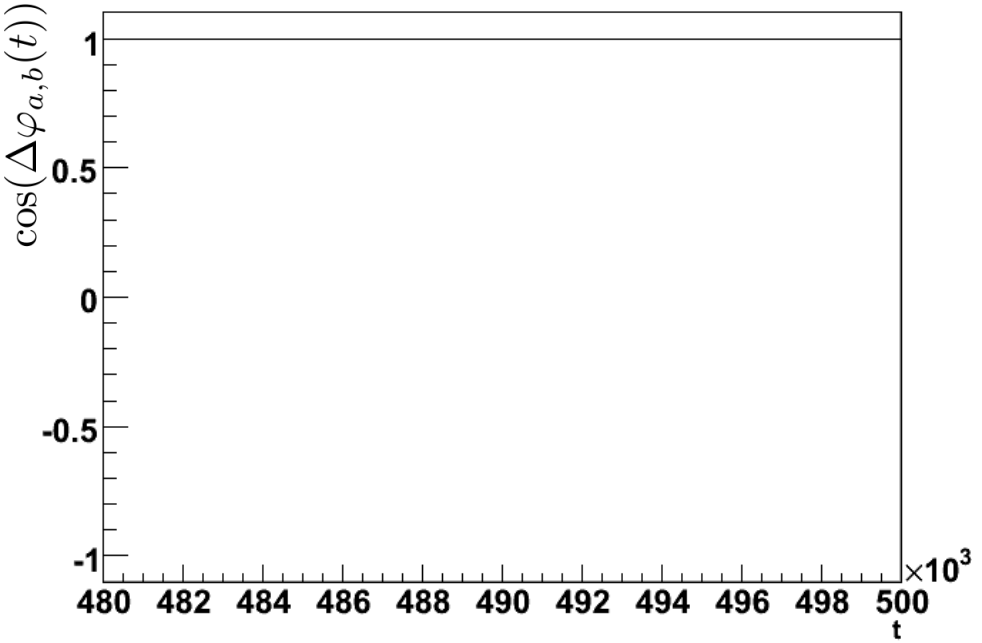} \\
\hline
\end{tabular}
\end{center}
\caption{Specimens of our experiments, where the $SC$ is verified. Column I: Billiard with the initial wave packet (\ref{egwp})  represented by the cloud. The arrow denotes the wave vector $\boldsymbol{k}$. Column II: Intensity (\ref{etini}). Column III: Portion of the plot of $\cos(\Delta\varphi_{a,b} (t))$ as function of time, where $\Delta\varphi_{a,b} (t)$ is the phase difference  of the wave function at the slits at time $t$. Time in unit of the time step size $\tau$. Position $x$ on the screen in unit of the space step size $\delta$.}
\lb{scs}
\end{figure*}
\begin{figure*}[p]
\begin{center}
\begin{tabular}{|c|c|c|}
\hline
Column I & Column II & Column III\\
\hline

c) \includegraphics[scale=0.171]{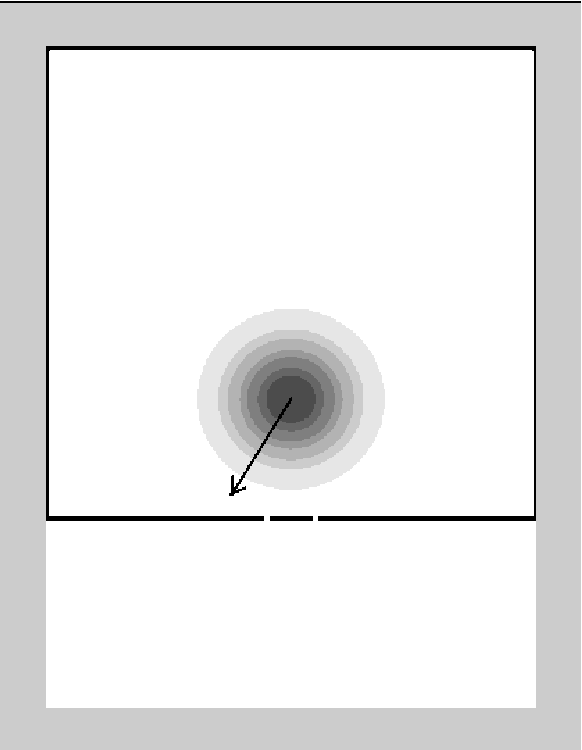} & \includegraphics[height=3.2cm]{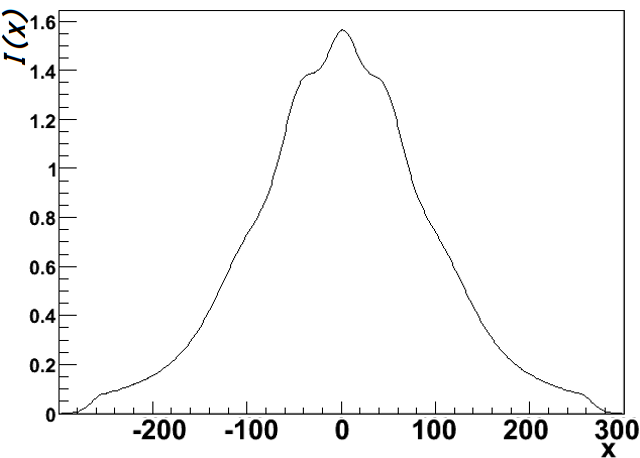} &       \includegraphics[height=3.2cm]{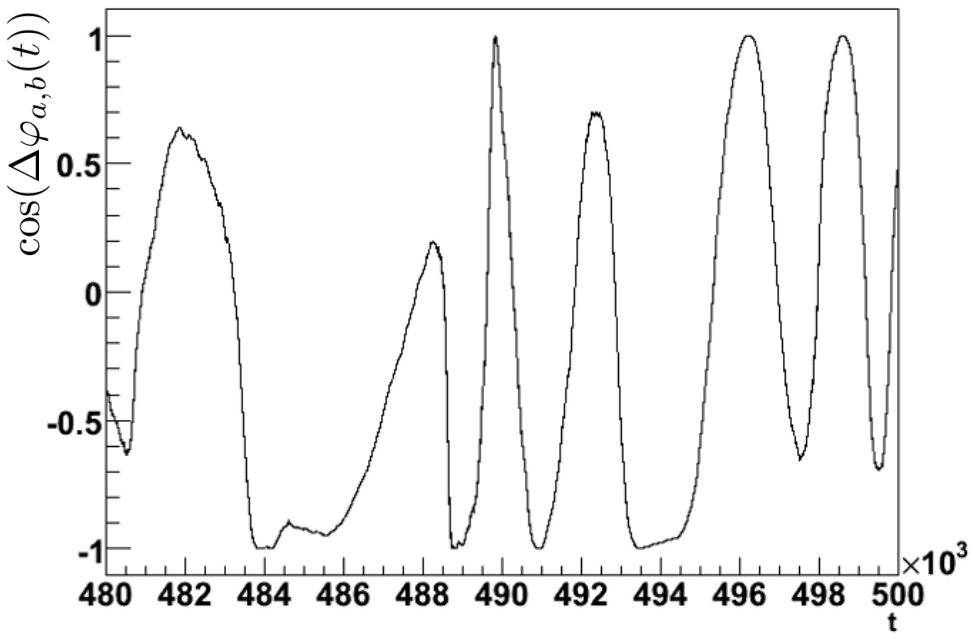} \\
\hline
d) \includegraphics[scale=0.171]{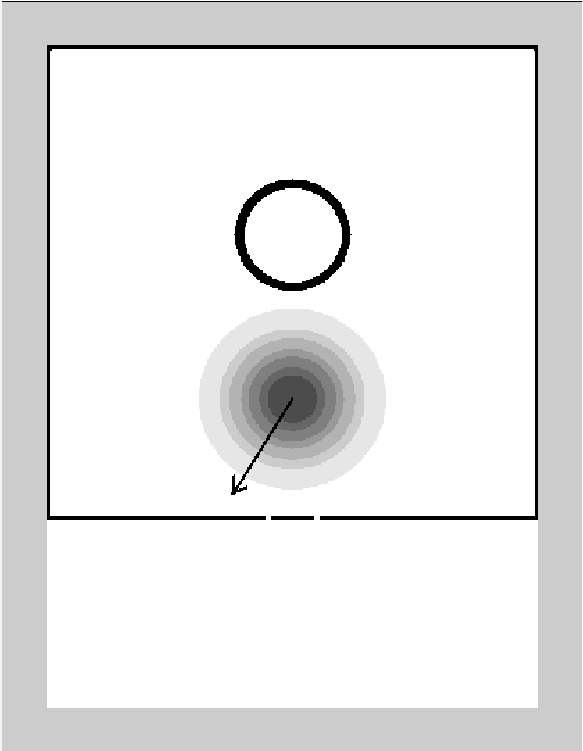} & \includegraphics[height=3.2cm]{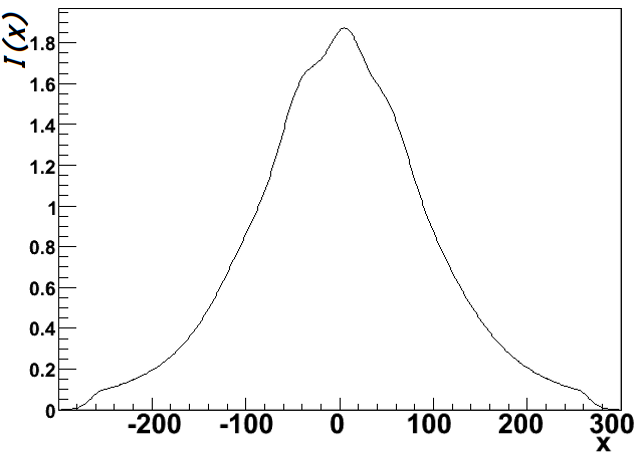} &         \includegraphics[height=3.2cm]{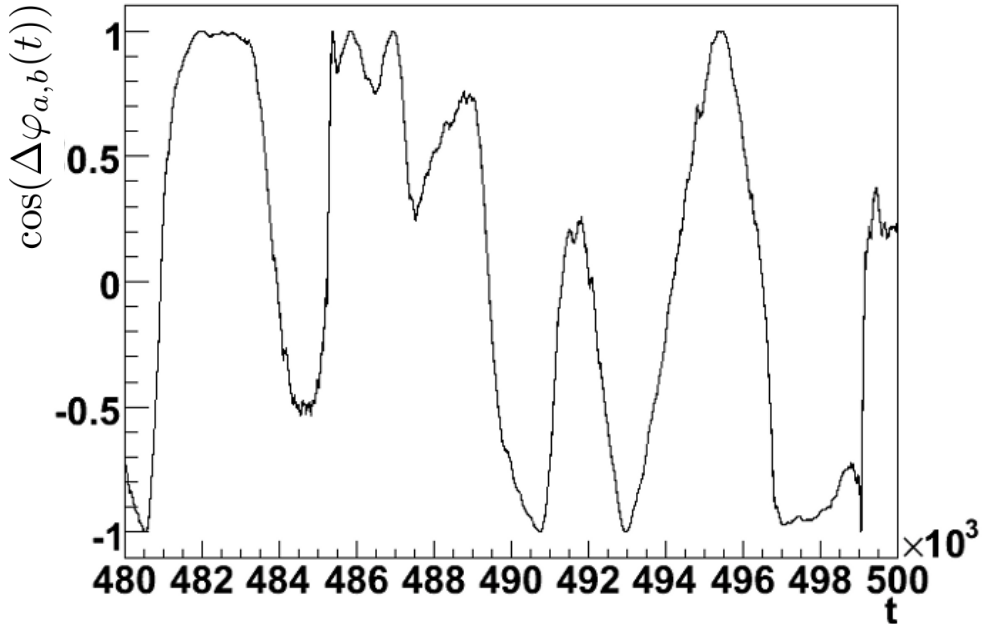} \\
\hline
e) \includegraphics[scale=0.171]{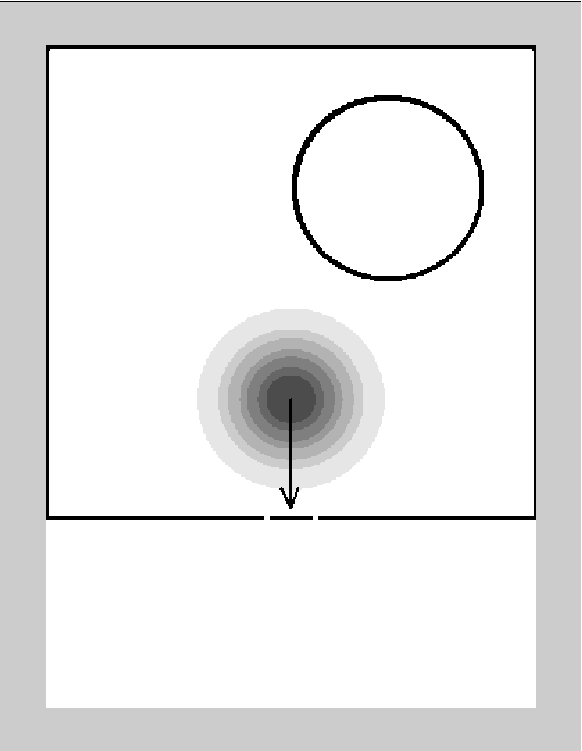} & \includegraphics[height=3.2cm]{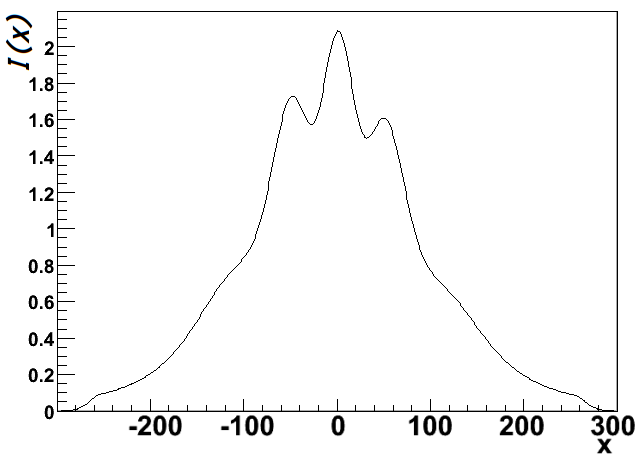} &       \includegraphics[height=3.2cm]{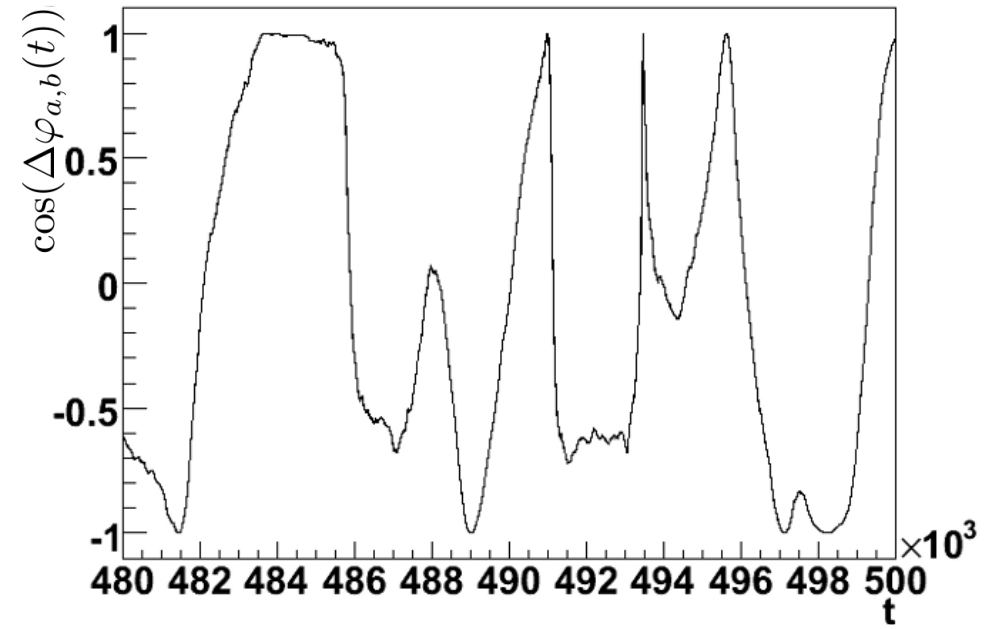} \\
\hline
f) \includegraphics[scale=0.171]{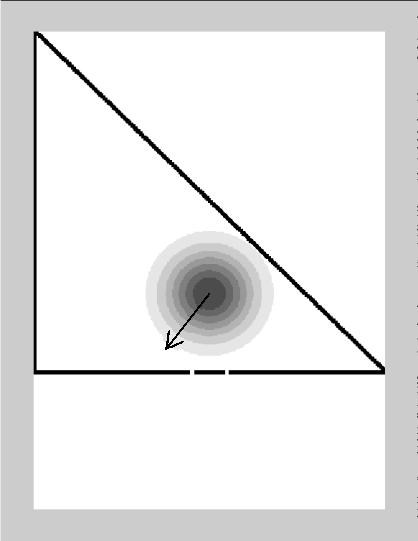} & \includegraphics[height=3.2cm]{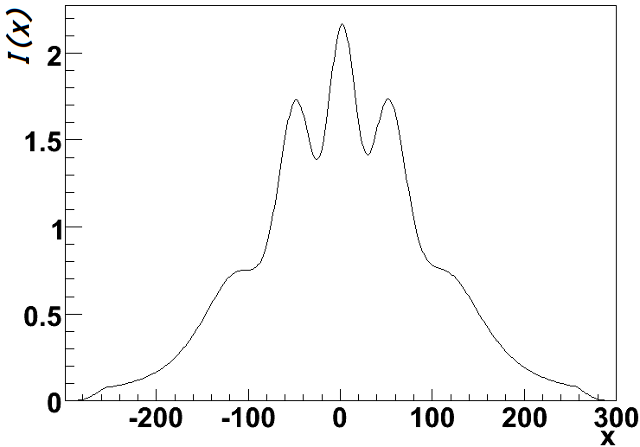} &         \includegraphics[height=3.2cm]{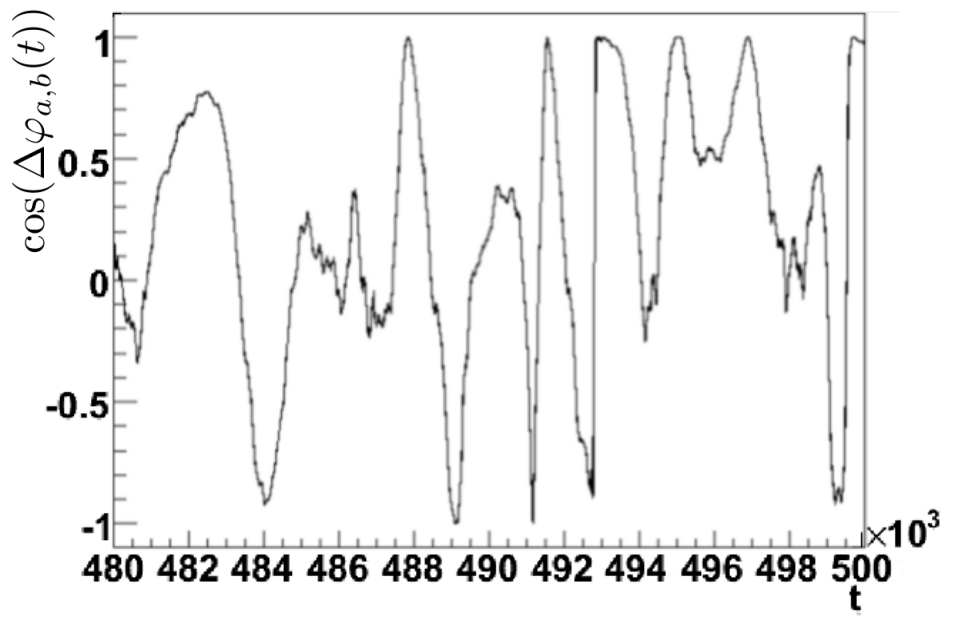} \\
\hline
\end{tabular}
\end{center}
\caption{Specimens of our experiments, where the $SC$ is violated. Columns I, II, III and units of $t$ and $x$ are the same as in Fig. \ref{scs}.}
\lb{scv}
\end{figure*}
\begin{figure*}
\begin{center}
\begin{tabular}{|c|c|c|}
\hline
Column I & Column II & Column III\\
\hline
\includegraphics[scale=0.171]{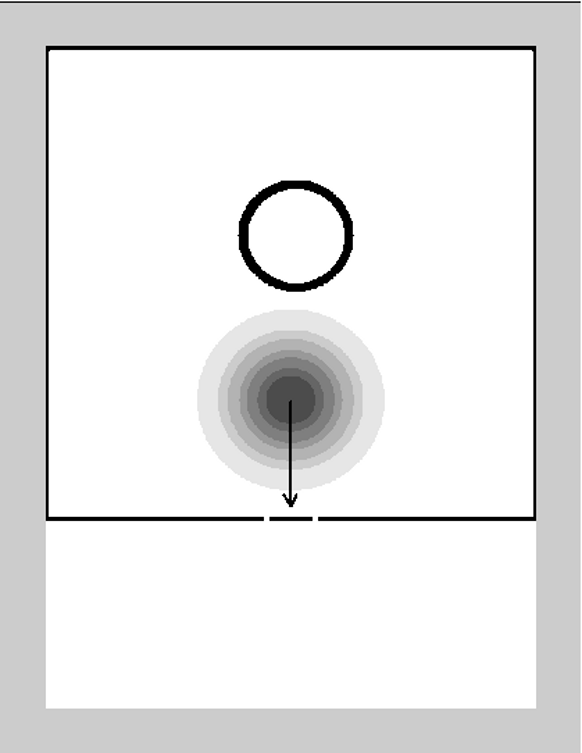} & \includegraphics[height=3.2cm]{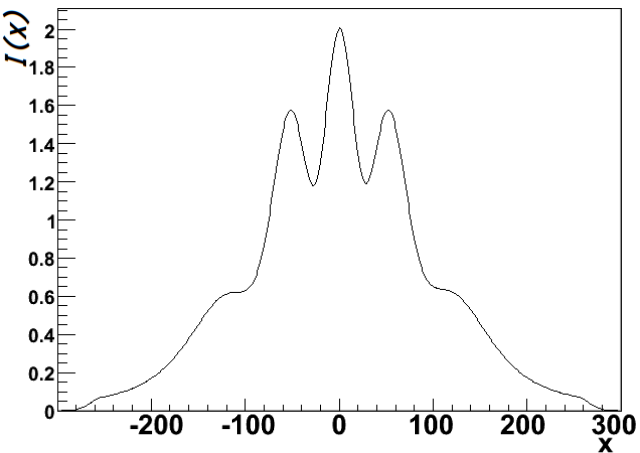} &         \includegraphics[height=3.2cm]{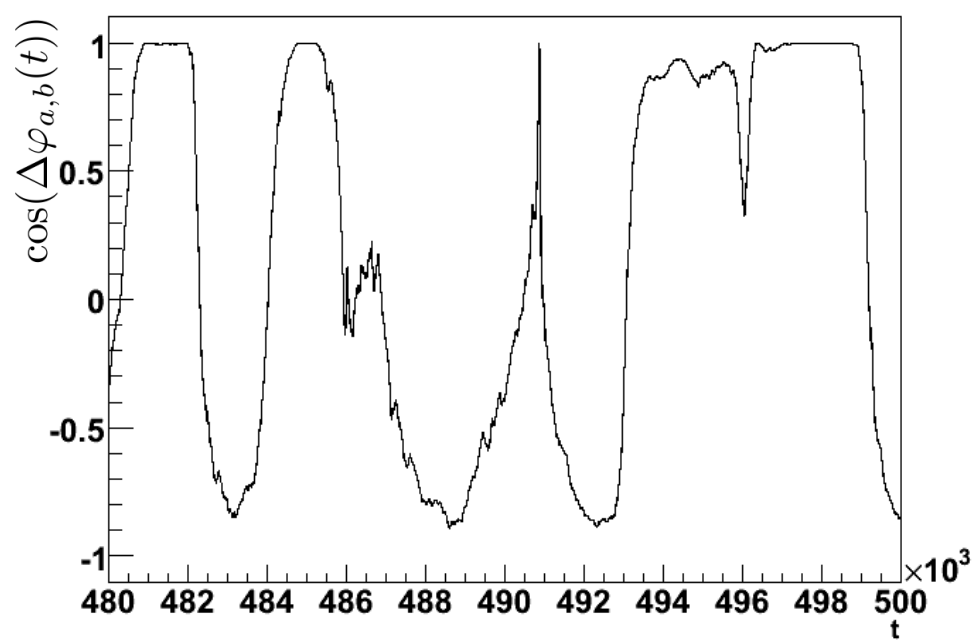} \\
\hline
\end{tabular}
\end{center}
\caption{Sensitivity of  interference patterns to the breaking of the $SC$.  Columns I, II, III and units of $t$ and $x$ are the same as in Fig. \ref{scs}. The billiard differs from the symmetric Sinai billiard $b)$ only for the coordinates of the center of the ring $(x=0, y=-0.6$ in $b)$ and $x=0.01, y=-0.6$ here).  This slight shift ($1/100$ of the linear dimension of the billiard) lowers the visibility of the fringes from $100$ \% to about 40 \% and causes the dephasing at the slits displayed in Column III.}
\lb{sip}
\end{figure*}

On considering this set-up (see Fig. 1), one expects that, if the billiard shape is symmetric with respect to the $y-$axis and the center $P_{0}$ and the wave vector $\boldsymbol {k}$ of the initial wave packet (\ref{egwp}) lie on this axis, the time evolution of (\ref{egwp})  should be exactly the same in each of the two halves of the billiard in which it is halved by the $y-$axis. Therefore in this case and irrespective of the classical  integrability of the billiards, one  should always find symmetrical interference patterns   with a fringe visibility of $100\%$. When on the contrary the symmetry just mentioned is not verified, such perfect interference patterns \footnote{Henceforth, by the expression ``perfect interference patterns'' we will mean interference patterns symmetric w.r.t the $y$-axis and with a fringe visibility $(I_{max}-I_{min})/I_{max}$ equal to $100 \%$.}\lb{f} should never occur, because the time evolution of (\ref{egwp}) should be different in the two halves  of the billiard. We now make  rigorous these intuitive considerations.\\

\bp
The solution of the problem {\rm (\ref{eptch})} has the property
\be\lb{ecs}
\psi(x,y;t)=\psi(-x,y;t), \q (x,y) \in \mathbb{R^{\rm 2}}, \q t>0,
\ee
iff
\be\lb{escon}
\left\{ \begin{array}{ll}
V_{B}(x,y)=V_{B}(-x,y) \\
\psi_{0}(x,y)=\psi_{0}(-x,y),\;\q (x,y) \in \mathbb{R^{\rm 2}}.
  \end{array} \right.
\ee
\ep
{\bf Proof.}\ooos Performing the spatial reflection $x'=-x$, the Cauchy problem (\ref{eptch}) becomes
\be\lb{eptchh}
\left\{ \begin{array}{ll}
i \frac{\partial}{\partial t}\psi' (x',y;t)=[-\frac{1}{2}\triangle' + V'_{B}(x',y)]\psi'(x',y;t)\\
\psi'(x',y;0)=\psi'_{0}(x',y),
  \end{array} \right.
\ee
where
\bd
\triangle'=\frac{\partial^{2}}{\partial x'^{2}} + \frac{\partial^{2}}{\partial y^{2}},
\ed
\be\lb{evb}
V_{B}(x,y)=V_{B}(-x',y):=V'_{B}(x',y)
\ee
and
\be\lb{epsi}
\psi(x,y;t)=\psi(-x',y;t):=\psi'(x',y;t).
\ee
If (\ref{escon}) holds, the relationship (\ref{evb}) and  the the relationship (\ref{epsi}) at $t=0$ imply, respectively,
\bd
 V'_{B}(x',y)=V_{B}(x',y)\oos {\rm and}\oos \psi'_{0}(x',y)=\psi_{0}(x',y).
\ed
Consequently, the problem  (\ref{eptchh}) can be written as
\be\lb{eptchhh}
\left\{ \begin{array}{ll}
i \frac{\partial}{\partial t}\psi' (x',y;t)=[-\frac{1}{2}\triangle' + V_{B}(x',y)]\psi'(x',y;t)\\
\psi'(x',y;0)=\psi_{0}(x',y).
  \end{array} \right.
\ee
The problems (\ref{eptch}) and (\ref{eptchhh}) are formally identical, and thus $\psi(\cdot)$ and $\psi'(\cdot)$ are the same function. Hence recalling (\ref{epsi}),
\bd
\psi(x,y;t)=\psi(-x',y;t)=\psi'(x',y;t)=
\ed
\bd
=\psi(x',y;t)=\psi(-x,y;t), \q (x,y) \in \mathbb{R^{\rm 2}}, \q t >0.
\ed
If the condition (\ref{escon}) does not hold, the problems (\ref{eptch}) and (\ref{eptchh}) are different. Consequently,  $\psi(\cdot)$ and $\psi'(\cdot)$ are different functions. Recalling   again (\ref{epsi}), we can write
\bd
\psi(x,y;t)=\psi'(x',y;t)=\psi'(-x,y;t).
\ed
Since  $\psi(\cdot) \neq \psi'(\cdot)$, we  find now that, in general,
\bd
\psi(x,y;t) \neq \psi(-x,y;t),
\ed
 and thus (\ref{ecs}) cannot be satisfied.
\frsq

Henceforth, we shall call (\ref{escon}) {\em symmetry condition} ($SC$).\\
\bn\lb{no}
\brm
Recalling the expression of the initial state (\ref{egwp}), we find easily that the $SC$ is violated on the part of this state if at least one of the following two conditions holds:
 \bd
  k_{x} \neq 0;  x_{0}\neq 0.
 \ed
\ooos\ooos \ooos\ooos \ooos\ooos \ooos\ooos\ooos\ooos \ooos\ooos\ooos\ooos$\square$
\erm
\en

In  Fig(s). \ref{scs}, \ref{scv} we report a first  representative set of our outcomes  to highlight the role of the $SC$ in our experiments. In the experiments $a), b)$ the $SC$ is satisfied and thus (\ref{ecs}) holds. In particular, the phase difference $\Delta\varphi_{a,b} (t)$ of the wave function at the slits is equal to zero for all $t$. Therefore, we find perfect interference patterns, both when the billiard is regular (square billiard $a)$) and when it is chaotic (Sinai billiard $b)$). In the experiments in Fig. \ref{scv}, the $SC$ is violated from the the initial state in $c), d)$, from the billiard shape in $e)$ and both from the initial state and the billiard shape in $f)$. As we can see in Column III, an evident dephasing at the slits is always present and, irrespective of the billiard integrability, the fringes are either practically absent  (billiards $c)$, $d)$) or present  as traces of a different extent (billiards $e)$, $f)$). Notice that in the experiments $c), f)$ the dephasing at the slits can be ascribed only to the violation of the $SC$. Fig. \ref{sip}  shows how much the interference patterns are sensitive to the $SC$ violation.\\

\section{The role of the classical integrability of the billiards and of the initial state}
\ooos According to the results in the previous section, one might be tempted to say that our interference phenomena are determined by the $SC$ and that the classical integrability of the billiards has no role. In order to investigate this point and to understand better the effect of the $SC$ breaking on the part of the initial state (see Note \ref{no}), we have carried out  further numerical simulations shown in Fig. \ref{orb}, where we have considered a wider range of directions of $\boldsymbol{k}$ and of values for  $P_{0}$. In particular, we have chosen $P_{0}$ and  $\boldsymbol{k}$ so that   the classical orbit  matching  the initial state \footnote{That is the   orbit  pursued  in the billiard by a inertial  particle starting from  the center $P_{0}$ of (\ref{egwp}) with velocity equal to its wave vector $\boldsymbol{k}$.} comes out always periodic, whereas it is chaotic in $d),e)$ and periodic, but with a very long period, in  $c),f)$.
\begin{figure*}
\begin{center}
\begin{tabular}{|c|c|c|}
\hline
Column I & Column II & Column III\\
\hline
g) \includegraphics[scale=0.171]{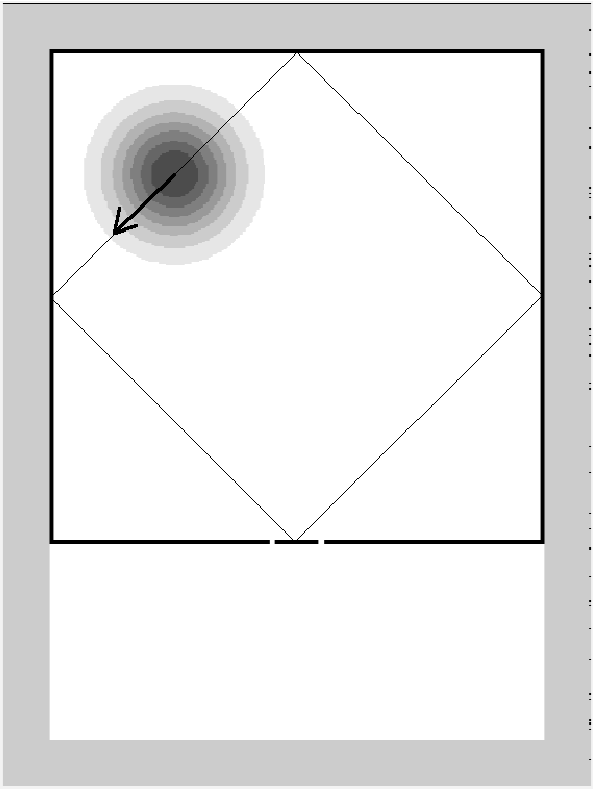} & \includegraphics[width=4.6cm]{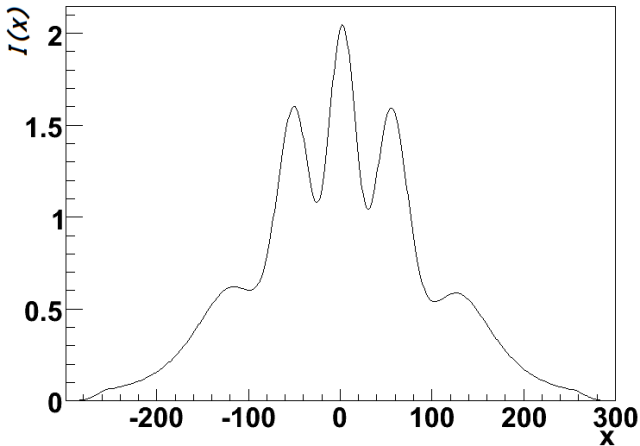} &       \includegraphics[width=4.9cm]{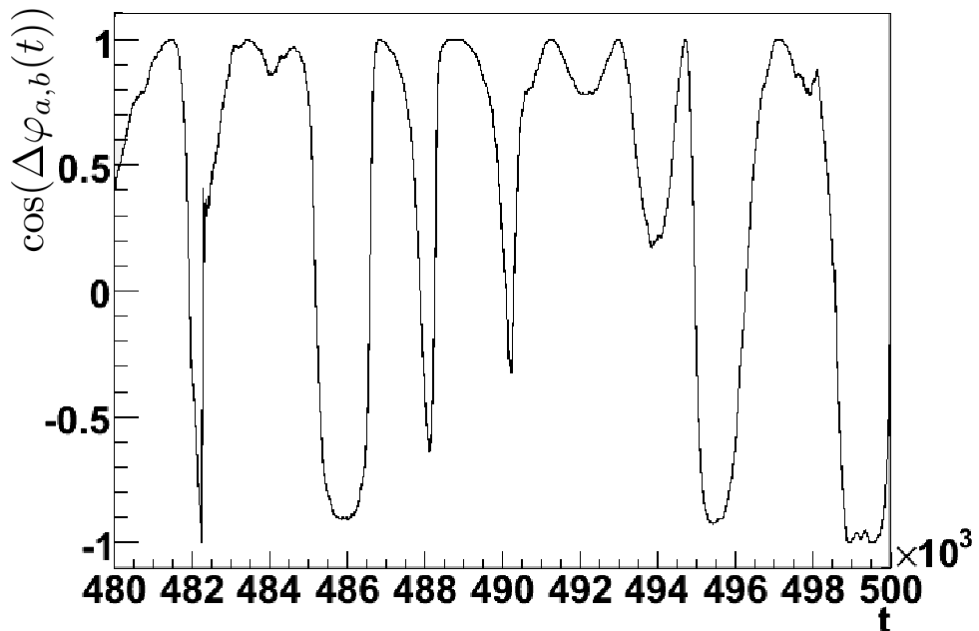} \\
\hline
h) \includegraphics[scale=0.171]{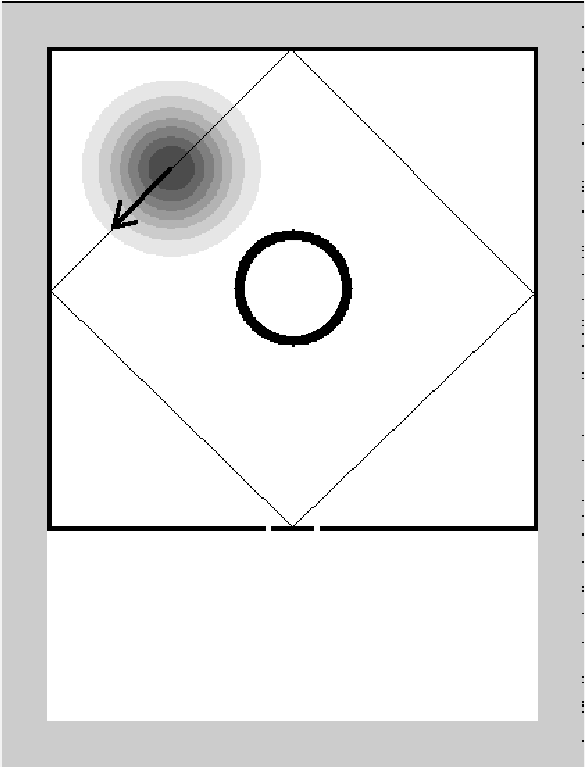} & \includegraphics[width=4.6cm]{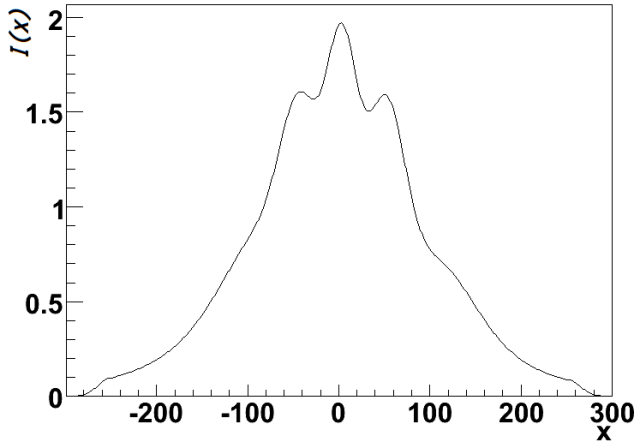} &       \includegraphics[width=4.9cm]{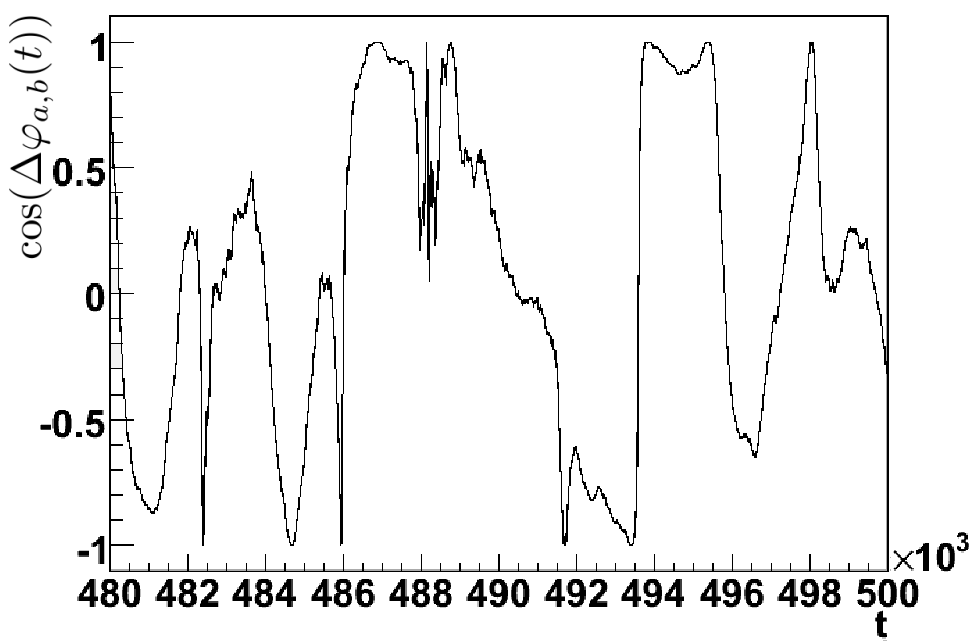} \\
\hline
i) \includegraphics[scale=0.171]{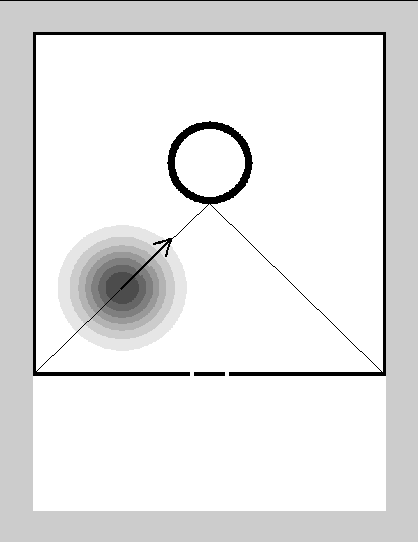} & \includegraphics[width=4.6cm]{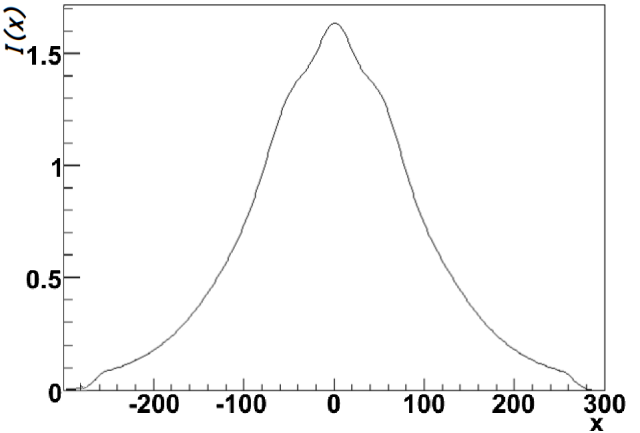} &       \includegraphics[width=4.9cm]{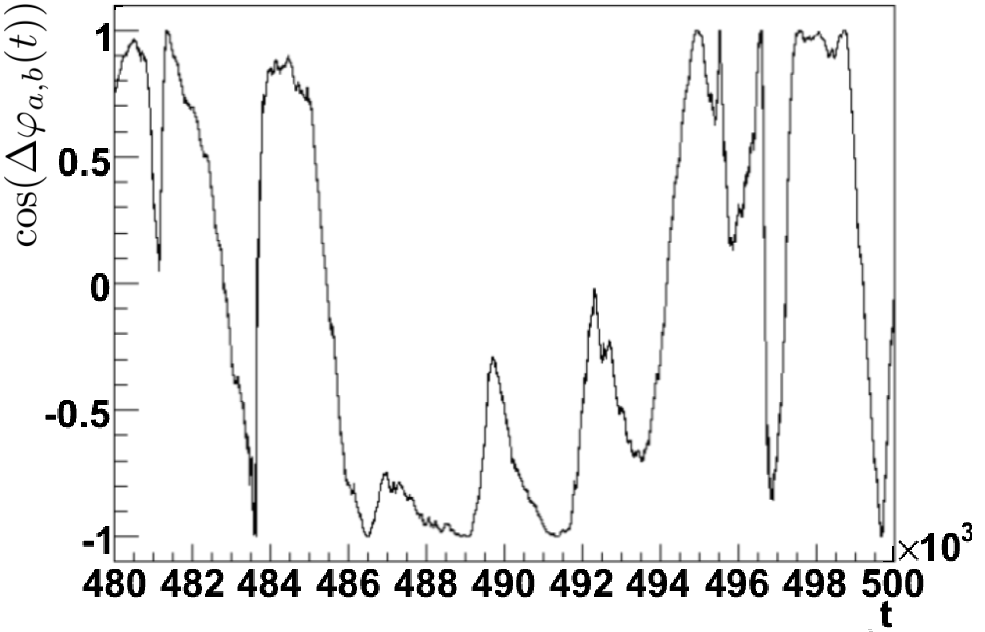} \\
\hline
j) \includegraphics[scale=0.171]{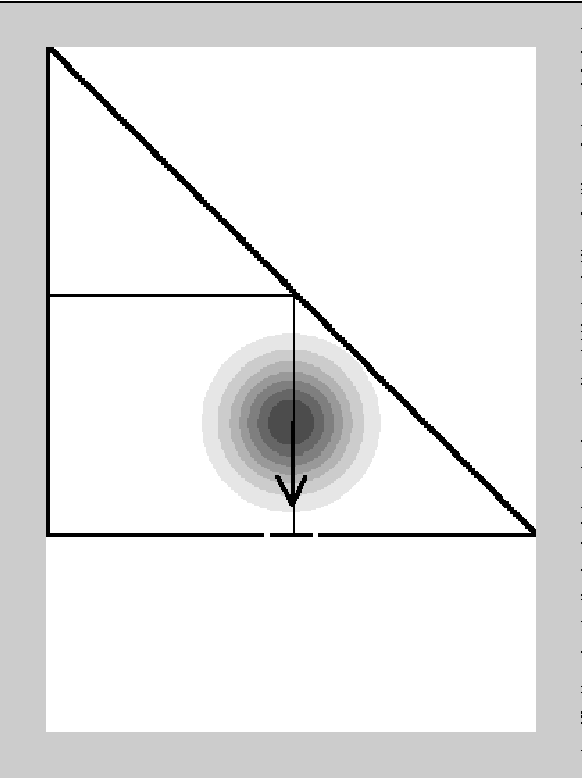} & \includegraphics[width=4.6cm]{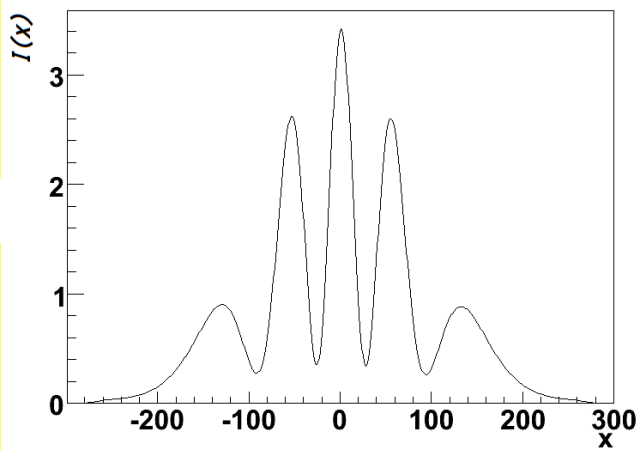} &       \includegraphics[width=4.9cm]{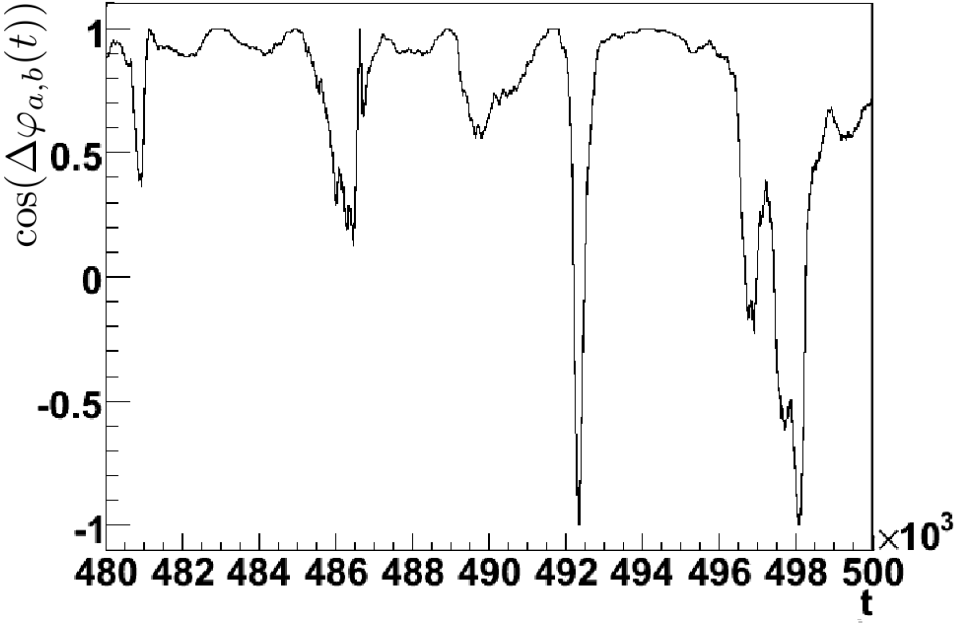} \\
\hline
k) \includegraphics[scale=0.171]{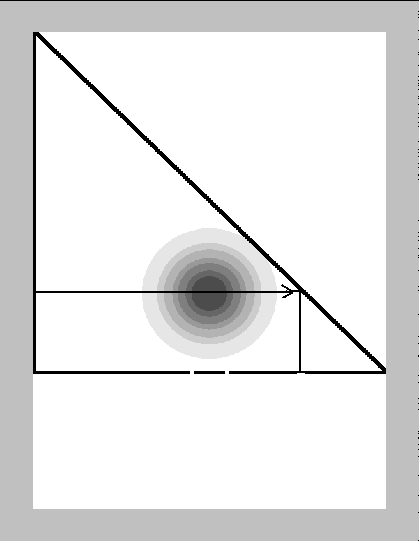} & \includegraphics[width=4.6cm]{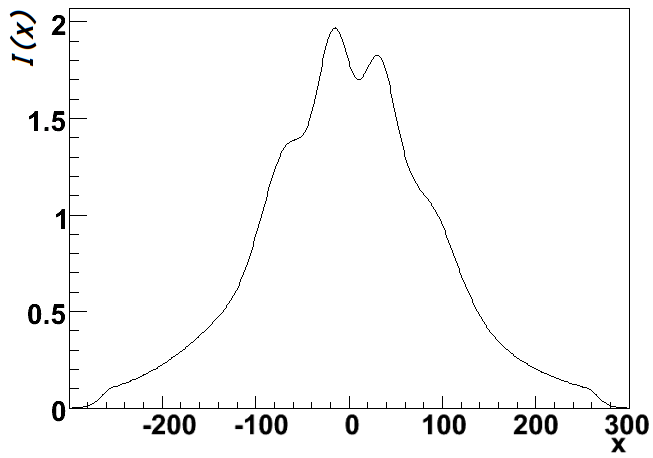} &       \includegraphics[width=4.9cm]{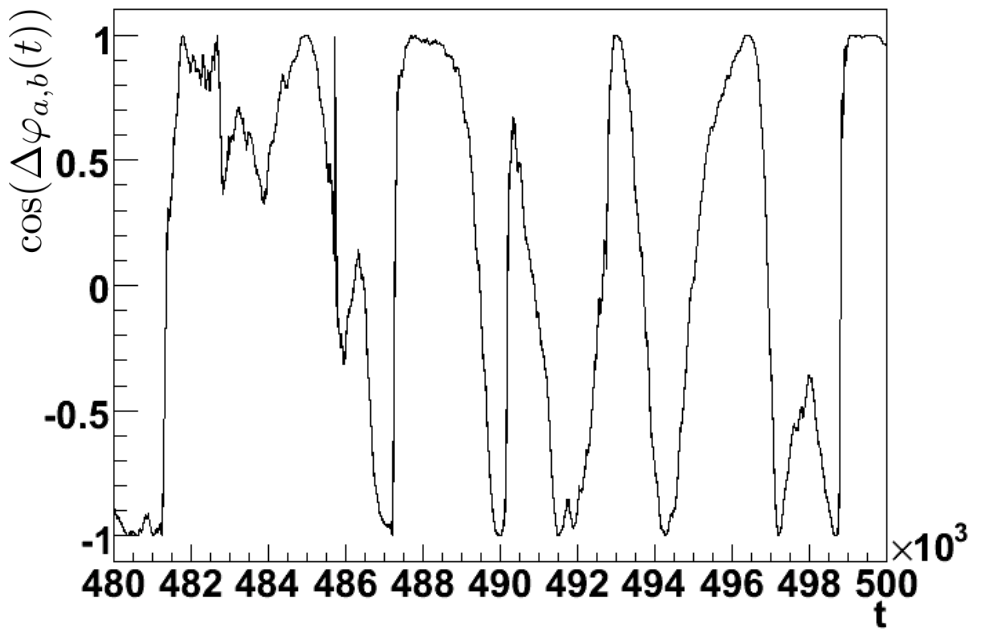} \\
\hline
\end{tabular}
\end{center}
\caption{Specimens of our experiments where, just as in Fig. \ref{scv}, the $SC$ is violated but where we have varied the center $P_{0}$ of the initial Gaussian wave packet  and the direction of its wave vector $\boldsymbol{k}$ so that the  classical orbit matching this state is periodic.  Columns I, II, III and units of $t$ and $x$  are the same as in Fig. 2. Inside each billiard is depicted  the corresponding classical orbit.}
\lb{orb}
\end{figure*}
 \begin{figure*}
\begin{center}
\begin{tabular}{|c|c|c|}
\hline
Column I & Column II & Column III\\
\hline
\includegraphics[scale=0.171]{SA.png} & \includegraphics[width=4.6cm]{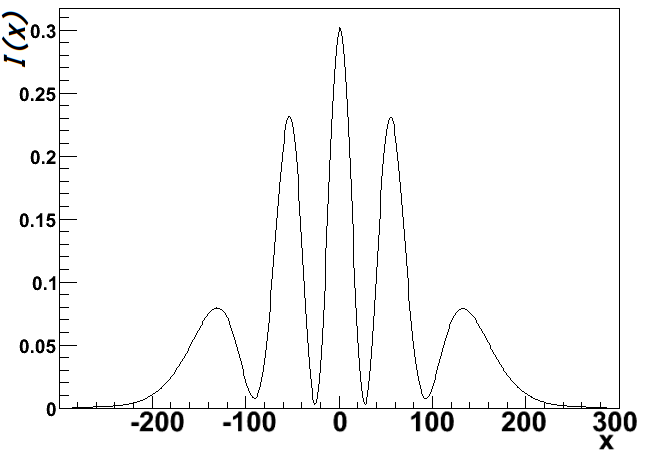} & \includegraphics[width=4.9cm]{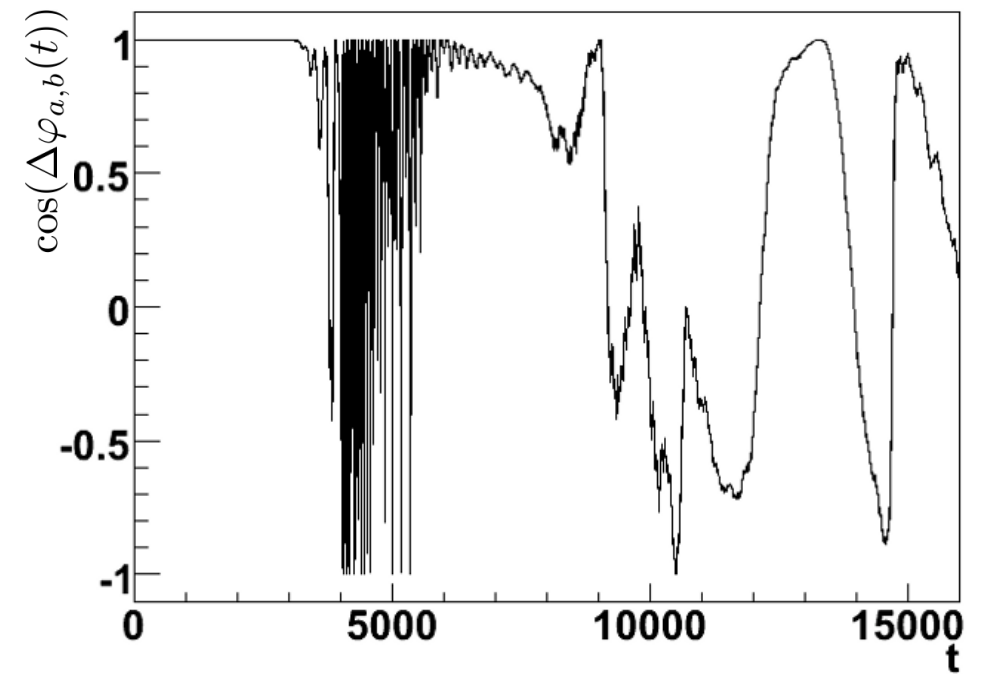}  \\
\hline
\end{tabular}
\end{center}
\caption{Example of a perfect $IIP$. The figure concerns the experiment $e)$ in Fig. \ref{scv}. Column I: Billiard. Column II: $IIP$  in the time interval $[0,10.000\tau]$. Column III: Plot of $\cos(\Delta_{a,b}\varphi (t))$. Units of $t$ and $x$ are the same as in Fig. 2.}
\lb{iip}
\end{figure*}

 Since  in  these new  experiments the $SC$ is always violated, we never find  perfect  interference patterns but, as we have already seen in Fig. \ref{scv}, we find only interference patterns variously perturbed. Examining all together  the experiments $c)\div k)$, we can see the following:

\bi
\item The visibility of the fringes  varies widely (it ranges from $0\%$  to $91.43\%$) and it can have very different values even in the same billiard  according to the classical orbit, {\em i.e.} according to $P_{0}$ and $\boldsymbol{k}$ (for example, see experiments $f), j), k)$).
\item   The visibility of the fringes is on average \footnote{The average is made on the results shown in Fig(s). \ref{scv}, \ref{orb}. These results are representative of a large number of experiments.}  more pronounced in the regular billiards ($36.02\%$ in these latter and $14.61\%$ in the chaotic billiards).
\item The experiment where the visibility of the fringes is close to $100\%$ concerns a regular billiard ($j)$).
\item In the experiments $g), h)$ the $SC$ is violated in the same way, but the  visibility of the fringes is higher in the regular billiard $g)$.
\ei

These facts  show that the features of the outcomes of our interference experiments are determined not only by the $SC$ breaking but also by the classical integrability of the billiards. We now explain  how this can happen by unveiling a likely physical mechanism  through which the $SC$ breaking, due to the values of $P_{0}$ and to the directions of $\boldsymbol{k}$, and the classical integrability of the billiards act.\\

 We begin by recalling  that the peculiarity of our experiments is that  the continuous interference of the two semicircular waves, coming out of the two slits, piles up on the viewing screen probability which gives rise to an  intensity pattern  whose  final appearance is  shown in our figures. At the beginning of the experiment, the initial wave packet (\ref{egwp}) is not yet dispersed  and  its wave fronts are perfectly regular (straight lines orthogonal to  $\boldsymbol{k}$). Thus, the corresponding interference pattern (initial  interference pattern ($IIP$)) may display  a good fringe   visibility  which, in some cases, is of $100\%$ (see Fig. \ref{iip}).
The percentage weight of this $IIP$ on the  visibility of the fringes in the final  intensity (\ref{etini}) depends on the following parameters:
   \bi
   \item [$p_{1})$] Ehrenfest's time (the time within which the initial  packet remains well localized and its centroid follows the corresponding classical trajectory).
   \item [$p_{2})$] The component $k_{y}(t)$ of the wave vector $\boldsymbol{k}$ when the wave packet impinges upon the billiard basis, in the early stages of the experiment.
   \item [$p_{3})$] The intersection point of the the classical orbit matching the initial state with the billiard basis.
    \ei

We now explain that the parameter $p_{1})$ determines the duration of the construction process of the $IIP$ and  that the parameters $p_{2}), p_{3})$ determine both the quantity of probability radiated from the slits, within Ehrenfest's time, and the fringe visibility and the symmetry of the $IIP$. As for $p_{1})$, the explanation is obvious: the longer   Ehrenfest's time is many times a well localized wave packet hits the slits with almost regular wave fronts. To explain the role of the parameters $p_{2}), p_{3})$, we note that, within Ehrenfest's time, the  wave packet can still be considered roughly Gaussian. In this case the component $j_{y}$ of the probability current at the slits and at  a given $t$ is
  \be\lb{ecabf}
  j_{y}(\al,0;t)=|\psi(\al,0;t)|^{2} k_{y}(t),
  \ee
   where $\al =-d/2,d/2$, being $d$ the distance between the slits. Recalling now  that the probability density in a Gaussian packet has a circular symmetry, we can see easily  that optimal values of $p_{3})$ imply large values of $|\psi(\al,0;t)|^{2}$ and  $|\psi(-d/2,0;t)|^{2} \simeq |\psi(d/2,0;t)|^{2}$. Thus by (\ref{ecabf}), the probability radiated becomes big and since the  slits emit equally strongly, the visibility of the fringes in the $IIP$ is high. Optimal values of $k_{y}(t)$ ({\em i.e.} $\boldsymbol{k}$ almost orthogonal to the billiard basis) imply, again by  (\ref{ecabf}), an increase in the probability radiation and a phase difference at the slits $\Delta_{a,b}\varphi (t)\simeq 0$. Thus  the $IIP$ turns out to be also symmetric w.r.t. the $y$-axis.\\

 Clarified the role of the parameters $p_{1})\div p_{3})$, we  make another consideration which we will apply soon and  which  reverses the reasoning path followed above. In virtue of (\ref{ecabf}), we can say that the more the probability radiated as $IIP$ is the bigger  $k_{y}(t)$   and  $|\psi(\al,0;t)|^{2}$ at the slits should be. For what we said above, this  means that the percentage weight of the $IIP$ on the  visibility of the fringes in the final  intensity  increases accordingly. \\
\begin{figure*}
\begin{center}
\begin{tabular}{|c|c|c|}
\hline
Column I & Column II & Column III\\
\hline
\includegraphics[height=6 cm]{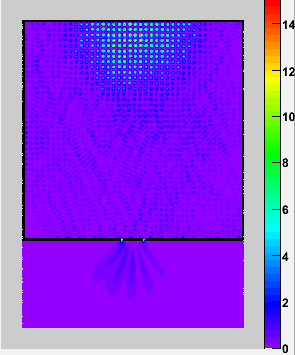} & \includegraphics[height=6 cm]{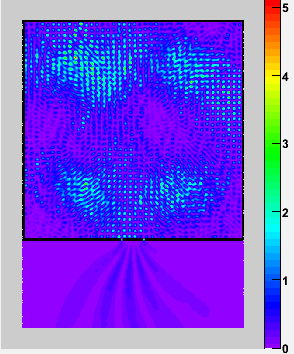} &    \includegraphics[height=6 cm]{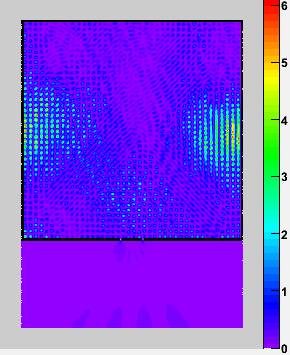} \\
\hline
\includegraphics[height=6 cm]{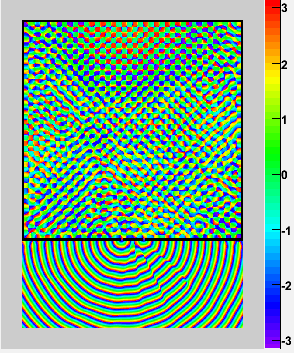} & \includegraphics[height=6 cm]{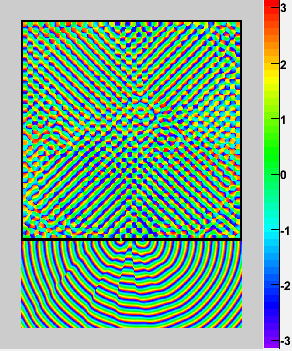} &    \includegraphics[height=6 cm]{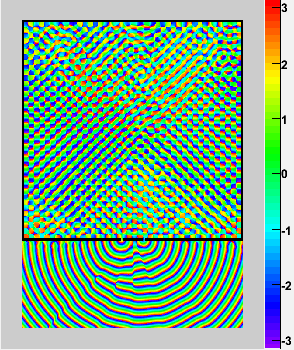} \\
\hline
\end{tabular}
\end{center}
\caption{Snapshots of recurrences. Top panels: Maps of  $\|\psi(x,y;t)\|^{2}$ showing  localizations of the initial wave packet along the classic orbit in the experiment $g)$ in Fig. \ref{orb}, at times $1.35\cdot 10^{6}\tau$ (Column I), $1.65\cdot 10^{6}\tau$ (Column II), $2.65\cdot 10^{6}\tau$ (Column III). On the right of each panel is the corresponding scale of values of  $\|\psi(x,y;t)\|^{2}$. Bottom panels:  Corresponding wave fronts, with on the right the scale of the values  of the phase of $\psi(x,y;t)$ in radians. The wave fronts inside the billiard look reminiscent of those in the initial state. Outside the billiard, we can see the two semicircular waves which interfere.}
\lb{wpl}
\end{figure*}
\begin{figure*}
\begin{center}
\begin{tabular}{|c|c|c|}
\hline
Column I & Column II & Column III\\
\hline
 \includegraphics[height=6 cm]{SSD.png} & \includegraphics[height=6 cm]{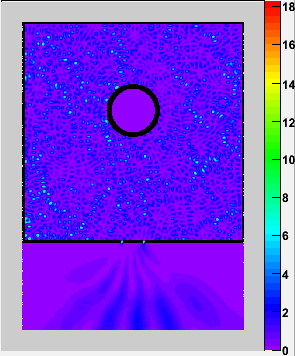}& \includegraphics[height=6 cm]{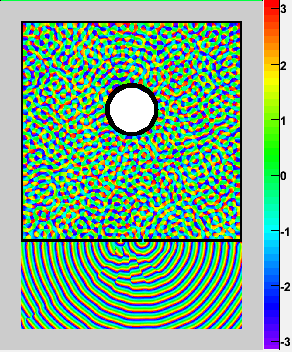}  \\
\hline
\end{tabular}
\end{center}
\caption{Typical example of lack of recurrences  in our experiments with chaotic billiards. Column I: Billiard $d)$.  Column II: Map of $\|\psi(x,y;t)\|^{2}$ at $t= 6 \cdot 10^{4} \tau$. Column III: Wave fronts of $\psi(x,y;t)$. On the right of the panels in Column II and in Column III are, respectively, the scale of the values of $\|\psi(x,y;t)\|^{2}$ and the scale of the values of the phase of $\psi(x,y;t)$ in radians. Notice that the wave fronts inside the billiard do not present any regularity.}
\lb{nwpl}
\end{figure*}

    There is another item that plays a crucial role in increasing  the  visibility of the fringes.  We will call it:
     \bi
     \item Recurrences of the initial wave packet.
     \ei
     To explain this item, we have to recall that the time evolution of a wave packet inside  closed square and triangle  billiards   shows revival phenomena \cite{rob}. This means that  the wave packet spreading, after a typical time (revival time), reverses itself and the wave packet relocalizes in the original shape or, after certain rational fraction of the revival time, in its smaller copies (fractional revivals). In the case of our billiards, the situation is  different, because  the  wave packet  continually loses probability from the slits and thus it can never return to its initial shape. Nevertheless, in the course of our experiments with regular billiards we have seen certain recurrences of the initial wave packet. These  manifest themselves especially in the wave fronts, which often look  reminescent of the regular ones in the initial wave packet.  Sometimes we have also noticed certain wave-packet localizations, resembling roughly to revivals or to fractional revivals (see Fig. \ref{wpl}). We believe that all these phenomena are the remains of  the revivals  in closed square and triangle billiards and that they are due to the complex interference processes that occur inside the billiards beyond  Ehrenfest's time. Now, as these recurrences repeat roughly features   of the initial wave packet, they may contribute to  the final intensity  with a pattern roughly similar to the $IIP$.  The extent of this contribution will depend on the percentage weight of the $IIP$ discussed above and on how the recurrences manifest themselves (frequency, duration, etc.). Things go differently in our Sinai billiards. The reason is that the revival phenomena are not, in general, expected \cite{tom} in chaotic billiards and, indeed,  we have not ever noticed  recurrences in the experiments $b),d),i)$  (see Fig. \ref{nwpl}), whereas in the experiments $e),h)$ we have  noticed only  weak appearances of them. So we have a first indication to say that the visibility of the fringes should be, in general, more relevant in regular billiards.
   \begin{table}[t]
\begin{center}
\begin{tabular}{|l|r|l|}
\hline
Regular Billiards & Probability& Visibility \\
\hline
$j)$     & $13.82 \%$ &$ 91.43\%$ \\
\hline
$f)$ & $4.80 \%$ & $34.09\%$ \\
\hline
$g)$    &  $4.74 \%$ & $53.66 \%$\\
\hline
 $c)$ & $3.31 \%$ & $\sim 0\%$ \\
\hline
$k)$ & $2.95 \%$ & $12.50\%$\\
\hline
 Sinai Billiards & Probability& Visibility \\
\hline
$e)$ & $7.34 \%$ & $30.95\%$ \\
\hline
$h)$ & $5.78\%$ & $22.50\%$\\
\hline
$d)$ & $2.60 \%$ & $0\%$\\
\hline
$i)$ & $2.26 \%$ & $ 0\%$\\
\hline
\end{tabular}
\end{center}
\caption{Correlation between the percentage of the total probability radiated as $IIP$  in the time interval $[0,50.000\tau]$ and the fringe visibility at end of the experiment. In the first column the billiard is denoted by the same symbol in Fig(s). \ref{scv}, \ref{orb}.}
\lb{tab2}
\end{table}
\begin{figure*}
\begin{center}
\begin{tabular}{|c|c|c|}
\hline
Column I & Column II & Column III\\
\hline
\includegraphics[scale=0.171]{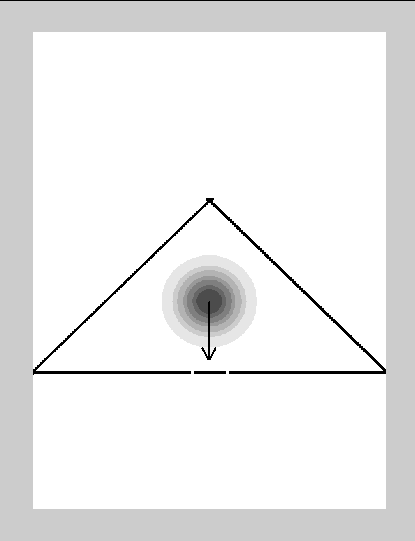} & \includegraphics[width=4.6cm]{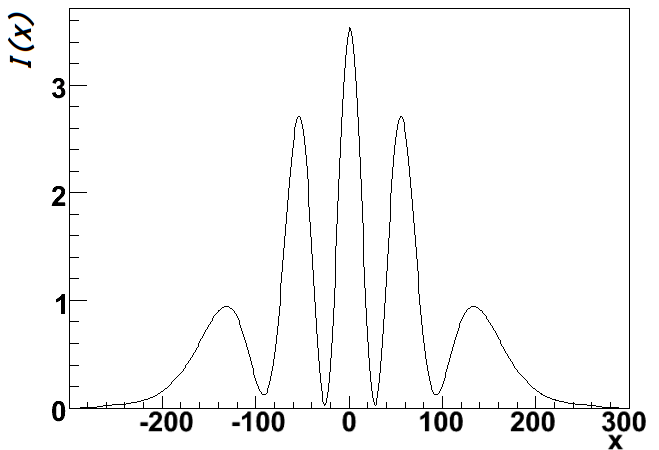} &         \includegraphics[width=4.9cm]{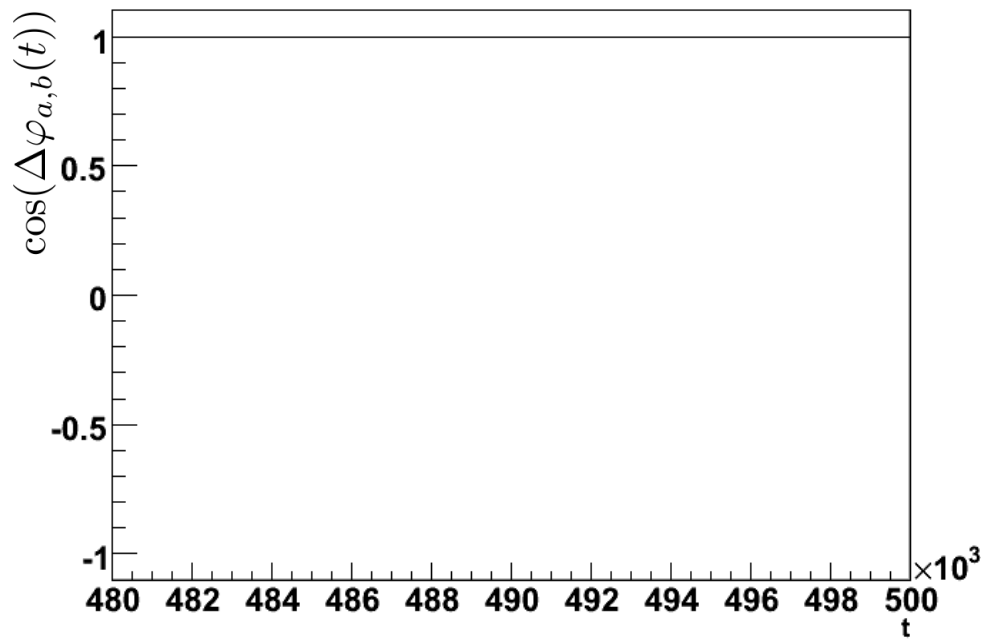} \\
\hline
\end{tabular}
\end{center}
\caption{Result relative to the triangle billiard in the unique set-up where the $SC$ is satisfied.  Columns I, II, III and units of $t$ and $x$ are the same as in Fig. 2.}
\lb{trs}
\end{figure*}
\begin{figure*}
\begin{center}
\begin{tabular}{|c|c|}
\hline
Column I & Column II \\
\hline
\includegraphics[scale=0.171]{SA.png} & \includegraphics[width=4.6cm]{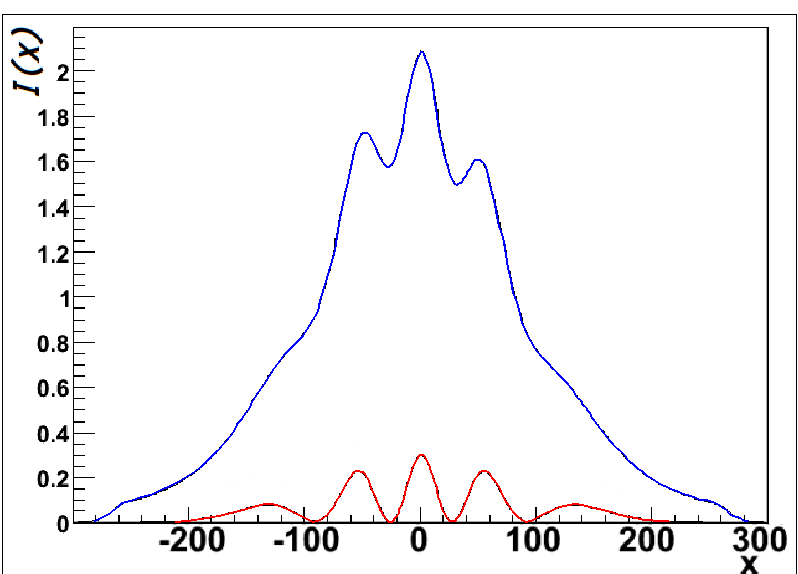} \\
\hline
\end{tabular}
\end{center}
\caption{Contribution of $IIP$  in Fig. \ref{iip}  to the final  intensity pattern in the experiment $e)$. Final intensity pattern  (blue curve). $IIP$ (red curve).}
\lb{ciip}
\end{figure*}

Another indication in favor of  regular billiards is that Ehrenfest's time in a quantum system is expected to be longer \cite{ter} when the corresponding classical system is regular. Indeed, we have verified that  Ehrenfest's time in our  regular billiards is about $22.000 \tau$, which is roughly three times longer than Ehrenfest's time in the chaotic billiards,  $b), e) ,i)$. In the billiards  $d),h)$, Ehrenfest's time is again about $22.000 \tau$, because the classical trajectory does not point directly to the billiard ring.\\

In conclusion, the mechanism based on subsequent contributions of the $IIP$ to the final intensity patterns depends on the parameters $p_{1})\div p_{3})$ in the early stages of the experiments and on the recurrences, which may occur at various times in the course of the experiments. The parameters $p_{1}),  p_{2})$ do not depend on the integrability of the billiards but they are at the same time related to the $SC$ violation from  the initial state (see Note \ref{no}) and to the classical orbit matching this state. The parameter $p_{1})$ and the recurrences depend on the  integrability of the billiards and thus they are at the origin of the differences  found in the experiments with the chaotic billiards.\\

 Coming back to  the  principle  previously hypothesized: ``the more  the probability radiated as $IIP$  the more  the percentage weight of  $IIP$ on the  visibility of the fringes in the final  intensity pattern'', we can now say that this principle works differently according to the billiard regularity. Indeed, the same percentage of probability emitted as $IIP$  is expected to give rise to a greater contribution to the fringe visibility in  regular billiards, owing to the recurrences. The principle  seems to be roughly confirmed in our experiments (see Table II).\\

  Actually, applying more specifically the mechanism of the subsequent contributions of the $IIP$, we can give a more detailed qualitative  explanation  for most of our experiments. We first explain  why  the experiment $j)$ gives rise to very clear interference fringes  (visibility of $91.43 \%$), notwithstanding the $SC$ is violated. Indeed, the  experiment $j)$ is the only one where  all three parameters $p_{1})\div p_{3}) $  occur in optimal manner and where, furthermore, the recurrences take place. As a consequence, the probability radiated as $IIP$  relative to the time interval $[0, 50.000\tau]$ is considerable  ($13.82  \%$ of the total probability) and the $IIP$ results in a perfect interference pattern. Intensity patterns roughly similar to this $IIP$  are then piled up in the final intensity pattern during the recurrences. It should be observed that the  statement in \cite{ca} that clear interference fringes occur only in regular billiards is based on an experiment  analogous to $j)$. However in $j)$, as well as in the experiment  in \cite{ca}, the $SC$  is violated. The presence of a dephasing at the slits due to this violation is well present (see the plot for $\Delta \varphi_{a,b}(t)$ in Column III of Fig. \ref{orb}). Consequently, the visibility of the fringes, as we have said, is lower than $100$ \%.  An interference pattern  with a visibility of the fringes of $100$ \%  it is instead obtained for this billiard in the set-up  in Fig.  \ref{trs}, the only one where the $SC$ is satisfied.\\

Unlike the experiment $j)$,  the parameters $p_{2}), p_{3})$ in the experiment $f)$ and the parameter $p_{3})$ in the experiment $k)$ are no longer optimal. As a consequence, the  probability radiated from the slits in the same time interval $[0, 50.000\tau]$ in $j)$ drops now, respectively, to  $4.8$ \% and to $2.95$ \% and the fringe visibility in the $IIP$ is lower. Thus, notwithstanding the recurrences, the fringe visibility in the final intensity drops as well, respectively, to $34.09 \%$  and to $12.50 \%$. In the experiment $i)$, none of the  parameters $p_{1})\div p_{3})$ is optimal and furthermore  it absent any recurrence phenomenon. As a consequence, any interference trace in the final intensity patterns is practically washed away. The same  can be said for the experiment $d)$, even if in this case Ehrenfest's time is longer. In the experiments $g),h)$, the parameters $p_{2}), p_{3})$ are equal (the $SC$ is violated in the same way), $p_{3})$ is optimal and also $p_{1})$ is (recall that Ehrenfest's time in both billiards is the same). Therefore, the probability emitted  in time interval $[0, 50.000\tau]$ is  comparable,  its quantity is good (see Table \ref{tab2}) and also good is the fringe visibility in the $IIP$, but  the fringe visibility is higher in the regular billiard $g)$, because the recurrences  are more relevant in this billiard. In the experiment $e)$  the  parameters $p_{2}),p_{3})$ occur in optimal manner, thus  the probability, radiated as $IIP$ in the time interval $[0,10.000\tau]$ ({\em i.e.} practically in the first collision of the initial wave packet  with the  slits), turns out to be considerable ($5.05 \%$ of the total probability)  and it gives rise to  a perfect $IIP$  (see Fig. \ref{iip}). This is the largest contribution to the fringe visibility in the final intensity pattern, because  Ehrenfest's time is short  and  the phenomenon of the recurrences is very  weak. Indeed, subtracting this $IIP$ from the final  pattern  (see Fig. \ref{ciip}), one gets an intensity pattern quite comparable with those in the experiments $d),i)$. The experiment $c)$ is more difficult to explain, because  the fringe visibility in the corresponding $IIP$ is fairly good and although the probability emitted within  Ehrenfest's time is not much (see Table \ref{tab2}), one should expect all the same at least a few traces of interference in the final intensity pattern.\\

Even though our explanation for most of our experiments seems reasonable, it is always one of a qualitative nature and a more exhaustive explanation  would be advisable.

\section{Conclusions}

   \ooos In the present paper we have carried out an extensive  numerical simulation to verify the intriguing conclusion   in paper \cite{ca} that   classical chaos causes the disappearance of the interference fringes in experiments where  an initial  wave packet evolves inside a billiard domain, with two slits  on the boundary. Our investigation leads us to revise this conclusion. Indeed, we highlight another even more important factor which affects our interference phenomena: a spatial reflection symmetry concerning the experimental set-up (the $SC$ (\ref{escon})).\\

   When the $SC$ is verified, the time evolution of the initial state is identical in each of the two halves in which the billiard is halved by the $y$-axis (see Fig. \ref{setup}).  As a consequence, the phase difference at the slits is zero at any time and thus, at the end of each experiment and irrespective of the classical integrability of the billiards, we always find perfect interference patterns. When the $SC$  is not verified, the time evolution of the initial state  in the two  halves of billiard is different. Thus, some dephasing at the slits is always present, even in the regular billiards, where the dephasing can be (see experiment $c)$) on a level with that in the experiments with fully chaotic billiards. The scenario of the results becomes in this case rather complex. Irrespective of the billiard integrability, the intensity  patterns  show a great variability, ranging from cases where they are practically the sum of the intensities of two one-slit experiments to cases where they  are almost perfect interference patterns. It is just this scenario that  besides being correlated with  the violation of the $SC$ is also correlated with the classical integrability of the billiards. We have explained this correlation by calling in question a mechanism based on subsequent contributions of the $IIP$ (initial interference pattern) to the final intensity patterns. This mechanism depends both on the $SC$ violation from the initial state through the parameters $p_{1}), p_{2})$ and on the billiard integrability through the length of Ehrenfest's time (parameter $p_{1}))$ and the occurrence of the recurrences mentioned in the previous section.\\

In conclusion, the present paper presents two novelties. The first novelty is that we somewhat downgrade  the hypothesis that classical chaos is the main cause of the disappearance of the interference fringes. The second novelty is the mechanism of the $IIP$. Although this mechanism seems to give a convincing qualitative explanation  for most of our results, it cannot explain  in detail the features of each experiment. A more thorough   explanation would require a more detailed knowledge of it and a precise quantitative evaluation of the effects of each of the parameters $p_{1}) \div p_{3})$ and of the phenomenon of the recurrences. This latter is the most difficult to handle,  because  it stems from the complex  interference phenomena which occur inside the billiards in the course of the experiments and thus it may depend by the initial state, by the  billiard shape and, as we said, by the billiard integrability.

\begin{acknowledgments}
 This work has been supported by INFN, Sezione di Catania and by MIUR. We wish to thank Professor G. Casati and Professor A. Rapisarda for their interest in this work and for helpful discussions. We wish also to thank Professor L. Pappalardo for an important conversation and Doctor F. Caruso for drawing out our attention to papers \cite{ca}.

\end{acknowledgments}

\end{document}